\documentclass[twoside]{article}

\usepackage[accepted]{aistats2024}
\usepackage[ruled,vlined,linesnumbered]{algorithm2e}
\SetKwInput{KwInput}{Input}
\SetKwInput{KwOutput}{Output}
\usepackage{amsfonts,amsmath,,amssymb,amsthm}
\usepackage{mathtools}
\usepackage{mathrsfs}
\newenvironment{outline}{\begin{proof}[Proof outline]}{\end{proof}}

\usepackage{bm}
\usepackage{booktabs} 
\usepackage{longtable}
\usepackage{tabu}

\usepackage{caption} 
\usepackage{subcaption}
\usepackage{comment}
\usepackage[dvipsnames]{xcolor}

\usepackage{derivative}
\usepackage{dsfont}

\usepackage[shortlabels]{enumitem}
\usepackage{empheq}

\usepackage{float}
\usepackage{fontawesome}
\usepackage{pifont}

\usepackage{graphicx}

\usepackage[colorlinks,linkcolor=black,citecolor=black,urlcolor=black,bookmarks=true,hypertexnames=false]{hyperref}
\usepackage{cleveref}

\usepackage{latexsym}
\usepackage{lipsum}

\usepackage{marvosym}
\usepackage{multirow} 

\usepackage{pgfplotstable}
\pgfplotsset{compat=1.18}
\usepackage{pgfmath, pgffor}
\usepackage{tikz}
\usetikzlibrary{calc}
\usetikzlibrary{positioning}
\usetikzlibrary{shapes,decorations}
\usetikzlibrary{decorations.pathreplacing,calligraphy}

\usepackage{stmaryrd}

\usepackage[most,breakable]{tcolorbox}
\usepackage{thmtools}
\allowdisplaybreaks
\usepackage{thm-restate}
\usepackage[textsize=tiny]{todonotes}
\usepackage{twemojis}

\usepackage{upgreek}

\usepackage{wasysym}
\usepackage{wrapfig}

\usepackage{xargs} 
\usepackage{xfrac}
\usepackage{xspace}

\newcommandx{\todoy}[2][1=]{\todo[linecolor=OliveGreen,backgroundcolor=OliveGreen!25,bordercolor=OliveGreen,#1]{#2}}

\makeatletter
\newcommand*{\indep}{%
	\mathbin{%
		\mathpalette{\@indep}{}%
	}%
}
\newcommand*{\nindep}{%
	\mathbin{
		\mathpalette{\@indep}{\not}
	}%
}
\newcommand*{\@indep}[2]{%
	\sbox0{$#1\perp\m@th$}
	\sbox2{$#1=$}
	\sbox4{$#1\vcenter{}$}
	\rlap{\copy0}
	\dimen@=\dimexpr\ht2-\ht4-.2pt\relax
	\kern\dimen@{#2}%
	\kern\dimen@
	\copy0 
} 
\makeatother

\newtcbox{\mymath}[1][red]{%
    nobeforeafter, math upper, tcbox raise base,
    enhanced, colframe=#1!30!black,
    colback=#1!30, boxrule=1pt
}

\newcommand{\EE}{\mathds{E}} 
\newcommand{\RR}{\mathds{R}} 

\newcommand{\KL}{\ensuremath{\mathsf{KL}}\xspace}

\newcommand{\argmax}{\mathop{\mathrm{argmax}}}
\newcommand{\argmin}{\mathop{\mathrm{argmin}}}

\newcommand{\grad}{{\nabla}}

\usepackage{actuarialsymbol}

\def\restrict#1{\raise-.5ex\hbox{\ensuremath|}_{#1}}

\newcommand{\abs}[1]{\left|#1\right|}
\newcommand{\cbr}[1]{\left\{#1\right\}}

\ifx\BlackBox\undefined
\newcommand{\BlackBox}{\rule{1.5ex}{1.5ex}}  
\fi

\ifx\QED\undefined
\def\QED{~\rule[-1pt]{5pt}{5pt}\par\medskip}
\fi

\ifx\proof\undefined
\newenvironment{proof}{\par\noindent{\em Proof:\ }}{\hfill\BlackBox\\[.0mm]}
\fi

\ifx\theorem\undefined
\newtheorem{theorem}{Theorem}
\fi

\ifx\example\undefined
\newtheorem{example}{Example}
\fi

\ifx\property\undefined
\newtheorem{property}{Property}
\fi

\ifx\lemma\undefined
\newtheorem{lemma}{Lemma}
\fi

\ifx\proposition\undefined
\newtheorem{proposition}{Proposition}
\fi

\ifx\remark\undefined
\newtheorem{remark}{Remark}
\fi

\ifx\corollary\undefined
\newtheorem{corollary}{Corollary}
\fi

\ifx\definition\undefined
\newtheorem{definition}{Definition}
\fi

\ifx\conjecture\undefined
\newtheorem{conjecture}{Conjecture}
\fi

\ifx\axiom\undefined
\newtheorem{axiom}[theorem]{Axiom}
\fi

\ifx\claim\undefined
\newtheorem{claim}[theorem]{Claim}
\fi

\ifx\assumption\undefined
\newtheorem{assumption}{Assumption}
\fi

\newtheorem{lem}{Lemma}[section]

\newtheorem{defn}{Definition}[section]

\newtheorem{rem}{Remark}[section]

\newcommand{\abf}{\mathbf{a}}
\newcommand{\bbf}{\mathbf{b}}
\newcommand{\cbf}{\mathbf{c}}
\newcommand{\dbf}{\mathbf{d}}
\newcommand{\ebf}{\mathbf{e}}
\newcommand{\fbf}{\mathbf{f}}
\newcommand{\gbf}{\mathbf{g}}
\newcommand{\hbf}{\mathbf{h}}
\newcommand{\ibf}{\mathbf{i}}
\newcommand{\jbf}{\mathbf{j}}
\newcommand{\kbf}{\mathbf{k}}
\newcommand{\lbf}{\mathbf{l}}
\newcommand{\mbf}{\mathbf{m}}
\newcommand{\nbf}{\mathbf{n}}
\newcommand{\obf}{\mathbf{o}}
\newcommand{\pbf}{\mathbf{p}}
\newcommand{\qbf}{\mathbf{q}}
\newcommand{\rbf}{\mathbf{r}}
\newcommand{\sbf}{\mathbf{s}}
\newcommand{\tbf}{\mathbf{t}}
\newcommand{\ubf}{\mathbf{u}}
\newcommand{\vbf}{\mathbf{v}}
\newcommand{\wbf}{\mathbf{w}}
\newcommand{\xbf}{\mathbf{x}}
\newcommand{\ybf}{\mathbf{y}}
\newcommand{\zbf}{\mathbf{z}}

\newcommand{\Abf}{\mathbf{A}}
\newcommand{\Bbf}{\mathbf{B}}
\newcommand{\Cbf}{\mathbf{C}}
\newcommand{\Dbf}{\mathbf{D}}
\newcommand{\Ebf}{\mathbf{E}}
\newcommand{\Fbf}{\mathbf{F}}
\newcommand{\Gbf}{\mathbf{G}}
\newcommand{\Hbf}{\mathbf{H}}
\newcommand{\Ibf}{\mathbf{I}}
\newcommand{\Jbf}{\mathbf{J}}
\newcommand{\Kbf}{\mathbf{K}}
\newcommand{\Lbf}{\mathbf{L}}
\newcommand{\Mbf}{\mathbf{M}}
\newcommand{\Nbf}{\mathbf{N}}
\newcommand{\Obf}{\mathbf{O}}
\newcommand{\Pbf}{\mathbf{P}}
\newcommand{\Qbf}{\mathbf{Q}}
\newcommand{\Rbf}{\mathbf{R}}
\newcommand{\Sbf}{\mathbf{S}}
\newcommand{\Tbf}{\mathbf{T}}
\newcommand{\Ubf}{\mathbf{U}}
\newcommand{\Vbf}{\mathbf{V}}
\newcommand{\Wbf}{\mathbf{W}}
\newcommand{\Xbf}{\mathbf{X}}
\newcommand{\Ybf}{\mathbf{Y}}
\newcommand{\Zbf}{\mathbf{Z}}

\newcommand{\Ac}{\mathcal{A}}

\newcommand{\Dsf}{\mathsf{D}}

\newcommand{\Lsf}{\mathsf{L}}

\newcommand{\Nsf}{\mathsf{N}}

\renewcommand{\vec}[1]{\mathbf{\boldsymbol{#1}}}

\DeclareMathAlphabet{\mathsfit}{\encodingdefault}{\sfdefault}{m}{sl}
\SetMathAlphabet{\mathsfit}{bold}{\encodingdefault}{\sfdefault}{bx}{n}
\newcommand{\tens}[1]{\bm{\mathsfit{#1}}}

\DeclareMathSymbol{\Alpha}{\mathalpha}{operators}{"41}
\DeclareMathSymbol{\Beta}{\mathalpha}{operators}{"42}
\DeclareMathSymbol{\Epsilon}{\mathalpha}{operators}{"45}
\DeclareMathSymbol{\Zeta}{\mathalpha}{operators}{"5A}
\DeclareMathSymbol{\Eta}{\mathalpha}{operators}{"48}
\DeclareMathSymbol{\Iota}{\mathalpha}{operators}{"49}
\DeclareMathSymbol{\Kappa}{\mathalpha}{operators}{"4B}
\DeclareMathSymbol{\Mu}{\mathalpha}{operators}{"4D}
\DeclareMathSymbol{\Nu}{\mathalpha}{operators}{"4E}
\DeclareMathSymbol{\Omicron}{\mathalpha}{operators}{"4F}
\DeclareMathSymbol{\Rho}{\mathalpha}{operators}{"50}
\DeclareMathSymbol{\Tau}{\mathalpha}{operators}{"54}
\DeclareMathSymbol{\Chi}{\mathalpha}{operators}{"58}
\DeclareMathSymbol{\omicron}{\mathord}{letters}{"6F}
\usepackage[round]{natbib}
\renewcommand{\bibname}{References}

\usepackage{cleveref}
\allowdisplaybreaks

\begin{document}

%

%

\twocolumn[

\aistatstitle{Convergence to Nash Equilibrium and No-regret Guarantee in (Markov) Potential Games}

\aistatsauthor{ Jing Dong$^1$ \And Baoxiang Wang$^1$ \And  Yaoliang Yu$^2$ }

\aistatsaddress{ $^1$ Chinese University of Hong Kong, Shenzhen \And $^2$ University of Waterloo } ]

\begin{abstract}
In this work, we study potential games and Markov potential games under stochastic cost and bandit feedback. 
We propose a variant of the Frank-Wolfe algorithm with sufficient exploration and recursive gradient estimation, which provably converges to the Nash equilibrium while attaining sublinear regret for each individual player. 
Our algorithm simultaneously achieves a Nash regret and a regret bound of $O(T^{4/5})$ for potential games, which matches the best available result, without using additional projection steps. 
Through carefully balancing the reuse of past samples and exploration of new samples, we then extend the results to Markov potential games and improve the best available Nash regret from $O(T^{5/6})$ to $O(T^{4/5})$. 
Moreover, our algorithm requires no knowledge of the game, such as the distribution mismatch coefficient, which provides more flexibility in its practical implementation. 
Experimental results corroborate our theoretical findings and underscore the practical effectiveness of our method. 

\end{abstract}

\section{Introduction}
One of the most popular solution concepts in multi-agent settings is Nash equilibrium, which originates from game theory to describe the behaviors of rational, selfish players \citep{roughgarden2010algorithmic}. It characterizes a stable state between the players, where no individual has any incentive to unilaterally deviate from their chosen strategy.  It is worth noting that the computational complexity of finding a Nash equilibrium in a general game is known to be PPAD-Hard \citep{chen2009settling,daskalakis2013complexity}. Nonetheless, it has been established that the computation of a Nash equilibrium becomes more feasible in specific game contexts, such as potential games \citep{monderer1996potential}, where a potential function is available to quantify how individual strategy changes affect collective utility. 

\begin{table*}[htb]
 \aboverulesep=0ex
 \belowrulesep=0ex
\caption{Summary of Nash regret bounds and regret for Markov potential game. Here, $n$ denotes the number of players, $m$ is the maximum size of action space among all players, $\max_{i\in[n]} |\mathcal{A}_i| = m$, $S$ is the size of the state space, $\kappa$ is the minimum stopping probability of the game, $D_\infty$ captures the distribution mismatch and $T$ is the time horizon. Note that when $S = 1$, the results are applicable to the potential game. }\label{table}
\begin{tabular}{l|c|c|c}
\toprule
\rule{0pt}{1.5EM}
& Nash regret                                            
& Regret                             
& Require Projection
\\ \midrule
\rule{0pt}{1.3EM}
\cite{giannou2022convergence}                      
& Asymptotic                                                  
& N/A                                  
& Yes              
\\ \midrule
\rule{0pt}{1.3EM}
\cite{zhang2021gradient}                           
& $\widetilde{O}\left(n^{5/6}\operatorname{poly}\left(\kappa^{-1},S, m\right)T^{5/6}\right)$ 
& N/A                                  
& Yes
\\ \midrule
\rule{0pt}{1.5EM}
\cite{leonardos2022global}
& $O \left(\frac{m^{5/6} D_\infty^{10/3} S^{5/3}T^{5/6}}{\kappa^{5/2}}\right)$
& N/A  
& Yes
\\ \midrule
\rule{0pt}{1.5EM}
\textbf{Ours}                                                        & \textbf{$O \left(\frac{ n^{3/2}mD_\infty T^{4/5}}{\kappa^{3}} + \frac{n^{5/8}m^{3/4}SD_\infty T^{4/5}}{\kappa^{8/3}}\right) $ }                                          
& \textbf{$O \left(\frac{mn^{3/2}S T^{4/5}}{\kappa^3}\right)$} 
& No \\\bottomrule
\end{tabular}
\end{table*}

In this paper, we study the potential game and its stochastic multi-step extension, the Markov potential game \citep{shapley1953stochastic,monderer1996potential}. We focus on the decentralized dynamic between the players under repeated games with bandit feedback. Specifically, over a time horizon, each player independently determines their strategy at each step and subsequently receives cost feedback based on their chosen strategies. Then they use this cost feedback information to refine their strategy. Under this decentralized dynamic, it is not clear what is the optimal way for a selfish player to choose their strategy. Thus, beyond the objective of arriving at a Nash equilibrium, a player should try to minimize the cost experienced while assuming the other players are malicious. One notion that evaluates the performance of an algorithm under possible malicious adversaries is the regret, which originates from the online learning community \citep{shalev2012online}. 

We are interested in designing learning algorithms that simultaneously enable the players to converge to a Nash equilibrium fast and with small regret. To measure the speed of convergence to Nash equilibrium, we use the notion of Nash regret, which is the cumulative gap to the Nash equilibrium. Designing such an algorithm confronts two main challenges. One is the nonstationarity for individual players, which arises as the environment changes when other players update their strategy, and consequently, the costs change. The other challenge is due to the bandit cost feedback. Since each player can only receive the cost feedback related to their chosen strategy, they must explore sufficiently alternative strategies to learn the Nash equilibrium. 

For potential games, previous works have shown that no-regret learning algorithms can converge to the equilibrium fast when additional information about the cost function is assumed \citep{panageas23} or when additional restrictions on the initial strategies of the players are imposed \citep{cohen2016exponentially,giannou2021rate,dong2023taming}. For the stochastic multi-step extension of the potential game, the Markov potential game, it remains uncertain whether such an algorithm can be designed. Prior attempts for the Markov potential game resort to the projected gradient descent algorithm, which either relies on a sophisticated analysis through proximal point or additional assumptions on the cost function to bound the gradient estimation error \citep{leonardos2022global,ding2022independent}. In both cases, the gradient estimation error can be large and make it difficult to balance fast convergence and exploration.

In this work, we introduce a variant of the Frank-Wolfe algorithm with exploration and recursive gradient estimation.
The use of a recursive gradient reduces the estimation error, enabling our algorithm to converge to the Nash equilibrium fast and ensuring each player has sublinear regret. 
Our algorithm can be extended to the Markov potential game and attains both Nash regret and regret of $O(T^{4/5})$, where $T$ is the length of the time horizon. Compared to previous decentralized algorithms for the Markov potential game, our algorithm improves the previous Nash regret of $O(T^{5/6})$ to $O(T^{4/5})$. Moreover, different from the previous algorithms, which are often based on projected gradient methods, our approach requires no additional projection step and therefore enjoys better computational complexity. Table \ref{table} summarizes the results for Nash regret and regret for (Markov) potential games. The algorithm is then evaluated empirically through a Markov congestion game task, which validates our theoretical findings and demonstrates the practical effectiveness of our method.

\section{Related Works}
\paragraph{Potential game and congestion game} Potential game and congestion game were introduced by \citet{monderer1996potential} and \citet{rosenthal1973class}, respectively. Notably, these two classes of games can be discussed interchangeably due to their equivalence. For potential (congestion) games, there is a long line of work that studies the dynamic of no-regret learning algorithms  \citep{cominetti2010payoff,palaiopanos2017multiplicative,heliou2017learning,daskalakis2011near,syrgkanis2015fast,chen2020hedging,hsieh2021adaptive}. In particular, under the bandit feedback model, \citet{heliou2017learning} showed the asymptotic last iterate convergence for the potential game and \citet{cominetti2010payoff,palaiopanos2017multiplicative} showed similar result for the congestion game. It has been shown that algorithms that are based on entropic regularization (e.g., exponential weights) can converge fast to the equilibria \citep{cohen2016exponentially,giannou2021rate,dong2023taming}, albeit with certain requirements on the initialization. 
Recent works \citep{cuilearning,panageas23} then established sublinear Nash regret for congestion game, with the latter additionally guaranteeing sublinear regret for each player. 

\paragraph{Markov potential game}
The study of Markov games, which is a stochastic multi-step extension of the one-step game, is initiated by \citet{shapley1953stochastic}. In stochastic optimal control, the study of the Markov potential game, which is the multi-step extension of the potential game, can be dated back to \citet{dechert2006stochastic,gonzalez2013discrete}. Recently there have been several works that establish the sample complexity of Nash equilibrium or Nash regret for the Markov potential game. \citet{song2021can} builds on the single agent algorithm Nash-VI and provides a $\Omega(\epsilon^{-2})$ sample complexity lower bound for finding $\epsilon$-approximate Nash equilibrium. They then introduced Nash coordinate ascent and established a $O(\epsilon^{-3})$ upper bound. However, their algorithm requires some coordination between the players. When each player updates their strategies independently and the exact gradient information is available (which may require knowledge of the cost function), \citet{chen2022convergence,zhang2022global,guo2023markov,sun2023provably} showed a sample complexity bound of $O(\epsilon^{-2})$ based on variations of gradient descent algorithms, which can be translated to $O(\sqrt{T})$ Nash regret. When the gradient information can only be estimated through interacting with the other players and the environment, \citet{leonardos2022global,ding2022independent} gave sample complexity bounds of $O(\epsilon^{-6})$ (or $O(T^{5/6})$ Nash regret) and $O(\epsilon^{-5})$ (or $O(T^{4/5})$ Nash regret), respectively. 
However, both of their methods require solving a projection problem at each step as they are both based on the projected gradient descent. Moreover, the algorithm presented by \citet{ding2022independent} relies on a regression subroutine and assumptions on cost structures, which is not realizable in the tabular Markov congestion game. 

We summarize the existing algorithms when no exact gradient information is available in Table \ref{table}.

\section{No-regret Learning for Potential Games}
In this section, we discuss the formulation of potential games and our algorithm for no-regret learning. Some of the techniques are used in the next section, where we generalize the results to Markov potential games.

\subsection{Potential Games}

Let $\Gamma$ be a strategic game with a finite number of players, $\mathcal{N}=\{1,2, \ldots, n\}$. Each player has a finite set of strategies, and we denote the strategy set of player $i$ as $\mathcal{A}_i$. The cost function associated with player $i$ is denoted by 
$c_i: \mathcal{A} \rightarrow [0,1]$, where $\mathcal{A}=\mathcal{A}_1 \times \mathcal{A}_2 \times \cdots \times \mathcal{A}_n$ denotes the set of joint strategy profiles. Denote $m=\max_{i \in [n]}|\mathcal{A}_i|$. Given a joint strategy $a \in \mathcal{A}$, we write $a = (a_i, a_{-i})$, where $a_i$ is the strategy of the $i$-th player and $a_{-i}$ is the strategy of all other players. We refer to the deterministic strategy profile introduced above as a pure strategy. Alternatively, each player $i$ can also play a randomized (mixed) strategy $\pi_i$ according to playing probability distributions over their strategy set $\Delta(\mathcal{A}_i)$. Upon playing a mixed strategy $\pi = \pi_1 \times \cdots \times  \pi_n$, the expected cost incurred for player $i$ is $\EE_{a_i \sim \pi_i, a_{-i} \sim \pi_{-i}}\left[c_i(a_i, a_{-i})\right] = \langle\pi_i, c_i(\cdot, \pi_{-i}) \rangle =: c_i(\pi_i, \pi_{-i})$, where $c_i(\cdot, \pi_{-i}) = \EE_{a_{-i} \sim \pi_{-i}} [c_i(\cdot, a_{-i})]$. In the stochastic setting, player $i$ receives a sample $C_i(a)$ from a distribution with mean $c_i(a_i, a_{-i}) $.

In potential games, $\Gamma$ admits a potential function $\Phi: \mathcal{A} \rightarrow \mathbb{R}$ such that for all $a_i, a_i^\prime \in \mathcal{A}_i$, $a_{-i} \in \mathcal{A}_{-i}$,
\begin{align}\label{eq:potential}
    c_{i} (a_i, a_{-i}) - c_{i} (a_i^\prime, a_{-i}) 
    = \Phi(a_i, a_{-i}) - \Phi(a_i^\prime, a_{-i}) \,.
\end{align}
We slightly abuse the notations and denote $\Phi(\pi) = \EE_{a \sim \pi} [\Phi(a)]$. By the definition of the potential function, we have $\nabla_{\pi_i} c_i(\pi) = c_i(\cdot, \pi_{-i}) = \nabla_{\pi_i} \Phi(\pi)$,
and $c_{i} (\pi_i, \pi_{-i}) - c_{i} (\pi_i^\prime, \pi_{-i}) = \Phi(\pi_i, \pi_{-i}) - \Phi(\pi_i^\prime, \pi_{-i})$. 
Clearly, every global minimum of $\Phi$ is a Nash equilibrium. As such, there always exists a pure strategy Nash equilibrium in a potential game. 

Noticing that $\nabla_{\pi}^2 \Phi(\pi)$ is continuous, and followed by the compactness of $\Delta(\Ac)$, the potential function is smooth.
\begin{lem}[Smoothness]\label{lem:smooth}
    For any $\pi, \pi^\prime \in \Delta(\mathcal{A})$, there exists an $L$ such that
    $\|\nabla \Phi(\pi) - \nabla\Phi(\pi^\prime)\|_2 \leq L \| \pi - \pi^\prime\|_2 $.
\end{lem} 


\paragraph{Learning protocol} 
We study the potential game in the repeated game setting with bandit feedback. Throughout a time horizon $T$, at each time step $t \in [T]$, all players take a strategy $a^t$ based on their mixed strategy $\pi^t$. Subsequently, they receive some feedback on their chosen strategies. Then they use this cost information to update their strategy. With bandit feedback, when the players take action $a^t \sim \pi^t$, they observes $C_i(a_i^t, a_{-i}^t)$ only. 

\paragraph{Solution concepts}
A popular solution concept of the potential game is Nash equilibrium, which is defined as the following.
\begin{defn}[Nash equilibrium]
    A strategy $\pi^\ast = (\pi^\ast_1, \ldots, \pi^\ast_n)$ is called a \textbf{Nash equilibrium} if for all players $i \in [n]$, it holds $c_i(\pi_i^\ast, \pi_{-i}^\ast) \leq  c_i(\pi_i, \pi_{-i}^\ast) \,, \quad \forall \pi_i \in \Delta(\mathcal{A}_i)$.
\end{defn}

In learning algorithms, the agents progressively find the Nash equilibrium by improving the approximation to the Nash equilibrium. The approximate Nash equilibrium is defined as follows.
\begin{defn}[$\epsilon$-approximate Nash equilibrium]
    A strategy $\pi^\ast = (\pi^\ast_1, \ldots, \pi^\ast_n)$ is called an \textbf{$\epsilon$-approximate Nash equilibrium} if for all players $i \in [n]$, it holds 
    $c_i(\pi_i^\ast, \pi_{-i}^\ast) \leq  c_i(\pi_i, \pi_{-i}^\ast) + \epsilon\,, \quad \forall \pi_i \in \Delta(\mathcal{A}_i)$.
\end{defn}

Such approximate Nash equilibria can be obtained by achieving a sublinear Nash regret, which is defined below. 
\begin{defn}[Nash regret]
    Let $\{\pi^t\}^T_{t=1}$ be a sequence of mixed strategies played across a time horizon of $T$, the Nash regret after $T$ time step is defined as 
    \begin{align*}
        & \mathrm{Nash\text{-}Regret}(T) \\
        = \ & \sum^T_{t=1} \max_{i \in [n]} \left(c_i(\pi_i^t, \pi_{-i}^t) - \min_{\pi_i \in \Delta(\mathcal{A}_i)} c_i(\pi_i, \pi_{-i}^t) \right)\,.
    \end{align*}
\end{defn}

By finding a sublinear expected Nash regret, one can converge to the Nash equilibrium in the sense of \textit{best-iterate convergence}. This means there exists some $t \in [T]$, which $\pi^t$ forms an approximate equilibrium in expectation. If one further finds the high probability of Nash regret, then it also implies that almost all iterations are $\epsilon$-Nash equilibrium. We note that this concept of convergence is commonly used in the literature \citep{anagnostides2022last,leonardos2022global,ding2022independent,cuilearning,panageas23}.

From an individual player's perspective, it is unclear how a selfish player should update their strategy. Thus, a player should try to minimize the cost experienced while assuming the worst costs possible with malicious adversaries. The performance of an online algorithm under malicious adversaries is captured by the notion of regret. We refer to the algorithms that achieve sublinear regret as no-regret algorithms. 
\begin{defn}[Regret of the $i$-th player]
    For an arbitrary sequence of $\{\pi_{-i}^t\}_{t=1}^{T=1}$, the regret of an online algorithm is defined as 
    \begin{align*}
        \mathrm{Regret}(T) = \sum^T_{t=1} c_i(\pi_i^t, \pi_{-i}^t) -  \min_{\pi_i^\ast\in \Delta(\mathcal{A}_i)} \sum^T_{t=1} c_i(\pi_i^\ast, \pi_{-i}^t) \,.
    \end{align*}
\end{defn}

\subsection{Algorithm and analysis for Potential Games}

We first introduce a general algorithm we proposed for potential games.
Our algorithm is largely based on the stochastic Frank-Wolfe method with a recursive gradient estimator \citep{ZhangSMHK20}, but with a carefully tuned exploration step to achieve sublinear regret. At each step $t \in [T]$, player $i$ plays a mixed strategy $\tilde{\pi}^t$, which is formed by mixing $\pi^t$ with a uniformly random distribution $s_i$ at a probability of $\mu$ (Line 3). Subsequently, upon receiving the instantaneous cost $C_i^t$, player $i$ uses the important sampling method to construct a one-sample gradient estimate of $\nabla{\pi_i} \Phi(\tilde{\pi}^t)$ (Line 5). This estimate is then used to create a recursive gradient estimator (Line 6), which is then used in the subsequent Frank-Wolfe update steps (Line 7 and 8).

\begin{algorithm}
\DontPrintSemicolon
\KwInput{Parameters $\eta_t$, $\rho_t$, $\mu$.}
Initialize $\pi_i^1$, $d_i^0$, for $i \in [n]$\;
\For{$t = 1, \ldots, T$}{
    $\tilde{\pi}_i^{t} = (1 - \mu)\pi_i^{t} + \mu s_i$, where $s_i$ is a uniformly random distribution on $\Delta(\mathcal{A}_i)$\;
    Play $a_i^t \sim \tilde{\pi}_i^t$ and receive cost $C^t_i$\;
    Let $(\hat{C}^t_i)_{a_i}= \frac{C^t_i \cdot \mathbb{I} \{a_i = a_i^t\}}{(\tilde{\pi}_i^t)_{a_i}} \,, \quad \forall a_i \in \mathcal{A}_i $\;
    Update $d^t_i = (1-\rho_t)d^{t-1}_i + \rho_t \hat{C}_i^t$\;
    Update $\hat{\pi}_i^{t+1} = \argmin_{\pi_i \in \Delta(\mathcal{A}_i)} \langle \pi_i, d^t_i \rangle$\;
    Update $\pi_i^{t+1} = (1-\eta_t)\pi_i^t + \eta_t \hat{\pi}_i^{t+1}$\;
}
\caption{Stochastic Frank-Wolfe algorithm with exploration for potential game}\label{alg}
\end{algorithm}

Previous efforts have explored the application of the Frank-Wolfe method to games \citep{cuilearning}. An inherent advantage of the Frank-Wolfe method lies in its ability to address the Frank-Wolfe gap, defined as $G(\pi^t) = \max_\pi (\pi^t - \pi)^\top \nabla \Phi(\pi^t)$. It minimizes this gap for each step, while a small cumulative Frank-Wolfe gap implies a small Nash regret, because 
\begin{align*}
    & \max_{\pi_i^\prime \in \Delta(\mathcal{A}_i)} \Phi\left(\pi_i^t, \pi_{-i}^t\right) -  \Phi\left(\pi_i^\prime, \pi_{-i}^t\right) \\
    = \ & \max_{\pi_i^\prime \in \Delta(\mathcal{A}_i)}\left(\pi^t_i - \pi_i^\prime\right)^\top \nabla_{\pi_i}\Phi\left(\pi_i^t, \pi_{-i}^t\right)  \\
    \leq  \ &  \max_{\pi^\prime}\left(\pi^t - \pi^\prime\right)^\top \nabla\Phi\left(\pi^t\right) 
    =  G(\pi^t) \,.
\end{align*}

In the case of bandits feedback, analyzing the Frank-Wolfe method typically requires bounding the gradient estimation error, i.e., $\|\nabla \Phi(\pi^t) - \hat{\nabla} \Phi(\pi^t) \|_2$. One of the most straightforward methods for obtaining a gradient estimate is through the importance sampling method (Line 5). Yet this only provides an error bound of $\|\nabla \Phi(\pi^t) - \hat{\nabla} \Phi(\pi^t) \|_2 = O\left(\frac{1}{\sqrt{\mu}}\right)$, where $\mu$ is the exploration parameter. The error would have accumulated to $O\left(\frac{T}{\sqrt{\mu}}\right)$ and resulted in large Nash regret. To overcome this, \cite{cuilearning} uses the regression method step to estimate the gradients, which results in an estimation error of roughly $\|\nabla \Phi(\pi^t) - \hat{\nabla} \Phi(\pi^t) \|_\infty = O(t^{-1/6})$ and a Nash regret of $O(T^{5/6})$. Our method instead uses a recursive gradient estimator $d^t$ (Lines 6 and 7) that reuses past gradient estimates. This enables us to enjoy a smaller estimation error of $\|\nabla \Phi(\pi^t) - d^t\|_2 = O\left(t^{-1/5}\right)$ and a subsequently smaller Nash regret of $O(T^{4/5})$. We also show that a $O(T^{4/5})$ regret can be attained simultaneously for each individual player. 

Another way to address the gradient estimation error is to use the projected gradient descent method, which only requires bounded gradient estimates $\hat{\nabla} \Phi(\pi^t)$ \citep{panageas23}. It has been shown that a $O(T^{4/5})$ Nash regret and $O(T^{4/5})$ regret for each individual player can be attained simultaneously.
However, when this method is extended to the Markov potential game, the Nash regret is only proved for $O(T^{5/6})$ and it is unclear whether individual players can enjoy sublinear regret simultaneously \citep{leonardos2022global}. It can also be shown that the utilization of a recursive gradient estimate, which may seem advantageous in its ability to reuse previous gradients for more informative updates, offers no improvement when used alongside the projected gradient descent method \citep{yuan2016influence}. Moreover, the projected gradient descent method requires an additional projection step at each iteration, which can be computationally expensive. 

The following theorem establishes the performance guarantee for Algorithm \ref{alg}. 
\begin{restatable}[Nash regret]{thm}{nashregret}
\label{thm:nash_regret}
If all players run Algorithm \ref{alg} with $\eta_t = \frac{1}{\sqrt{n}t^{4/5}}$, $\mu = \min\left\{\frac{1}{mn},\frac{m}{n^{1/8}T^{1/5}}\right\}$, $\rho_t = \frac{4\mu^{1/3}n^{1/3} m^{1/3}}{ t^\alpha }$, $T \geq mn^{7/8}$, then $\EE\left[ \mathrm{Nash\text{-}regret}(T) \right] = O \left(\left(m n^{15/8}L + n^{5/2}\right) T^{4/5}   \right)$.
\end{restatable}

\begin{rem}
    When $T \geq O \left(n^{75/8}m^{5}L^5\epsilon^{-5} \right)$, there exists $t \in \{1, \ldots, T\}$, such that $\pi^t$ is an $\epsilon$-approximate Nash equilibrium.
\end{rem}
\begin{rem}
    Using Markov inequality on Theorem \ref{thm:nash_regret}, it is immediate that the sublinear regret also holds with high probability. 
\end{rem}

Theorem \ref{thm:individual} investigates the case where one player chooses to play according to Algorithm \ref{alg} while other players play with arbitrary actions. It establishes an upper bound of the player's regret. 

\begin{restatable}[Regret for $i$-th player]{thm}{individualregret}\label{thm:individual}
Let $\{\tilde{\pi}_{-i}^t\}_{t=1}^T$ be an arbitrary sequence of strategies for all players except for player $i$. For player $i$, running Algorithm \ref{alg} with $\eta_t = \frac{1}{\sqrt{n}t^{4/5}}$, $\mu = \min\left\{\frac{1}{mn},\frac{m}{n^{1/8}T^{1/5}}\right\}$, $\rho_t =  \frac{4\mu^{1/3}n^{1/3} m^{1/3}}{t^{8/15}}$ and $T \geq mn^{7/8}$, we can obtains a regret at most 
\begin{align*}
    & \EE\left[\sum^T_{t=1} c_i(\tilde{\pi}_i^t, \tilde{\pi}_{-i}^t) -  \min_{\pi_i^\ast \in \Delta(\mathcal{A}_i)} \sum^T_{t=1} c_i(\pi_i^\ast, \tilde{\pi}_{-i}^t) \right]\\
    \le \ &  O \left(mnL T^{4/5}\right) \,.
\end{align*}
\end{restatable}

\paragraph{Extension to congestion game} We remark that both Theorem \ref{thm:nash_regret} and Theorem \ref{thm:individual} can be extended to congestion games, which is a potential game with exponentially many actions (each action is a combination of some resources). In such case, \citet{panageas23} introduced an algorithm that can simultaneously obtain a Nash regret and a regret of $O(\log(m)T^{4/5})$, under an additional condition of the feedback (the semi-bandit feedback model). To obtain an efficient algorithm with Nash regret and regret guarantees that are only logarithmically dependent on $m$ (which means polynomial in the number of resources), we can use the Carath\'{e}odory decomposition method to perform our algorithm on a polytope that can be described with $O (\log(m))$ fractional numbers. 
Additionally, under the bandit feedback model, we can estimate the gradients with existing estimators from the combinatorial bandit literature \citep{combes2015combinatorial}.
With these adjustments, our algorithm can obtain a regret of $O(\log(m)T^{4/5})$ for both Nash regret and regret for individual players. 

While we defer the proofs of both theorems to the appendix, we present an outline of the proof.

\begin{outline}[Theorem \ref{thm:nash_regret}]
By the definition of potential functions, we can first decompose the Nash regret into two parts, where the first part concerns the performance of the optimization algorithm, and the second part concerns regret caused by the exploration. That is,
\begin{align*}
    & \mathrm{Nash\text{-}regret(T)} \\ 
    \leq \ & \sum^n_{i=1} \sum^T_{t=1} \max_{\pi_i^\prime \in \Delta(\mathcal{A}_i)} \Phi\left(\tilde{\pi}_i^t, \tilde{\pi}_{-i}^t\right) -  \Phi\left(\pi_i^\prime, \tilde{\pi}_{-i}^t\right) \\
    \leq \ &  \max_{\pi_i^\prime} \Phi\left(\pi_i^t, \pi_{-i}^t\right) -  \Phi\left(\pi_i^\prime, \pi_{-i}^t\right) + 2n \mu \,,
\end{align*}
where the inequality is by the Lipschitzness of the potential function and the update rules of Algorithm \ref{alg}.

By the smoothness of the potential function, one can show the following descent inequality 
\begin{align*}
    & \Phi(\pi^{t+1}) \\
    \leq \ & \Phi(\pi^{t}) - \eta_t G(\pi^t) + 2\eta_t  \sqrt{n}  \left\|\nabla \Phi(\pi^t)  - d^t\right\|_2 + \frac{\eta_t^2 n L}{2} \,,
\end{align*}
where $G(\pi^t) = \max_\pi (\pi^t - \pi)^\top \nabla \Phi(\pi^t)$. 

By rearranging the terms and taking summation over $T$, we can upper bound $\sum^T_{t=1} G(\pi^t)$ as,
\begin{align*}
    & \sum^T_{t=1} G(\pi^t) \\
    \leq \ & \frac{2n}{\eta_{T+1}} + 2 \sqrt{n}  \sum^{T}_{t=1} \left\|\nabla \Phi(\pi^t)  - d^t\right\|_2 +  \frac{n L}{2}\sum^{T}_{t=1}\eta_t \,.
\end{align*}

Notice that this is the upper bound of the Nash regret, as $\max_{\pi_i^\prime \in \Delta(\mathcal{A}_i)} \Phi\left(\pi_i^t, \pi_{-i}^t\right) -  \Phi\left(\pi_i^\prime, \pi_{-i}^t\right) \leq G(\pi^t) $. 
It now only amounts to bounding the gradient estimation error $\left\|\nabla\Phi(\pi^t) - d^t\right\|_2$, which is derived in the following lemma through an induction argument.
\begin{restatable}[]{lem}{estdiff}\label{lem:est_diff}
Let $\eta_t =  \frac{1}{\sqrt{n}t^{3\alpha/2}} $, $\rho_t = \frac{4\mu^{1/3}n^{1/3} m^{1/3}}{ t^\alpha }$, $\mu \geq \frac{1}{mn}$, $\alpha, \beta \in (0,1)$ in Algorithm \ref{alg}. We have 
\begin{align*}
    \EE\left[\left\|\nabla\Phi(\pi^t) - d^t\right\|_2\right] \leq O \left(\mu \sqrt{n}L +  \frac{m^{4/3}n^{4/3}}{\mu^{1/3}(t+1)^{\alpha/2}}\right) \,.
\end{align*}

\end{restatable}
The theorem follows by selecting the appropriate parameters $\alpha$, $\beta$.
\end{outline}

\begin{outline}[Theorem \ref{thm:individual}]
Using the definition of potential functions and the update rules of Algorithm \ref{alg}, we can decompose the regret into two parts, where one is due to the exploration and one is due to suboptimal strategies, as
\begin{align*}
    & \sum^T_{t=1} c_i(\tilde{\pi}_i^t, \tilde{\pi}_{-i}^t) -  c_i(\pi_i^\ast, \tilde{\pi}_{-i}^t) \\
    \leq \ & \mu T + \left(\Phi(\pi_i^t, \tilde{\pi}_{-i}^t) -  \Phi(\pi_i^\ast, \tilde{\pi}_{-i}^t)\right) \,.
\end{align*}
Then, by the smoothness and linearity of $\Phi$, we obtain the following recurrence relationship 
\begin{align*}
    & \Phi(\pi_i^{t+1}, \tilde{\pi}_{-i}^t) -  \Phi(\pi_i^\ast, \tilde{\pi}_{-i}^t) \\
    \leq \ & (1 - \eta_t) \left( \Phi(\pi_i^{t}, \tilde{\pi}_{-i}^t) -  \Phi(\pi_i^\ast, \tilde{\pi}_{-i}^t)\right) \\
    & \ + 2 \eta_t\| \nabla_{\pi_i} \Phi(\pi_i^{t}, \tilde{\pi}_{-i}^t) - d_i^t\|_2 + \frac{\eta_t^2 L}{2} \,.
\end{align*}
Using proof by induction and by Lemma \ref{lem:est_diff}, we show that $\EE\left[ \Phi(\pi_i^{t}, \tilde{\pi}_{-i}^t) -  \Phi(\pi_i^\ast, \tilde{\pi}_{-i}^t)\right] = O(1/t^{1/5})$. Lastly, putting everything together yields Theorem \ref{thm:individual}.
\end{outline}

\section{No-regret Learning for Markov Potential Games}
In this section, we extend the algorithm and results to the Markov potential game, which is a natural stochastic extension of the potential game \citep{shapley1953stochastic,leonardos2022global}. 

In a Markov potential game, $n$ agents interact in a finite horizon MDP $\langle \mathcal{S}, (\mathcal{A}_i, c_i)_{i=1}^n, P, \mu_0, \kappa \rangle$. Here, $\mathcal{S}$ represents the state space, and for each player $i$, $\mathcal{A}_i$ and $c_i$ denote the strategy space and cost functions, respectively. We define the joint action space as $\mathcal{A} = \mathcal{A}_1 \times \cdots \times \mathcal{A}_n$, and the transition function is given by $P: \mathcal{S} \times \mathcal{A} \rightarrow \mathcal{S}$. This means that $P(\cdot \mid s,a)$ represents a probability measure over the state space $\mathcal{S}$ when the agents at state $s \in \mathcal{S}$ jointly take action $a \in \mathcal{A}$. Additionally, $\mu_0$ denotes the initial state distribution and $\kappa = \min_{s,a} \{\kappa_{s,a}\}$, where $\kappa_{s,a}$ is the probability that the game stops at state-action pair $(s,a)$. We assume $\kappa > 0$. 

For each player $i \in [n]$, a stochastic policy is denoted as $\pi_i \in \Delta(\mathcal{A}_i)^\mathcal{S}$, which is a mapping that associates a state $s \in \mathcal{S}$ with a probability distribution over actions in $\Delta(\mathcal{A}_i)$. Lastly, we define $m$ as the maximum cardinality among the strategy spaces of the individual players, i.e., $m = \max_{i\in[n]} |\mathcal{A}_i|$.

\paragraph{Learning protocol} The players learn their policy through $T$ episodes. Within each episode, at step $h$, all players observe the state $s^h$. Then, each player $i$ plays $a_i^h \sim \pi_i^t$ and receives a cost $C_i^h$ sampled from a distribution with mean $c_i(s^h, a_i^h, a_{-i}^h)$. Without loss of generality, we assume the costs $C_i^h$ are bounded between $[0,1]$. Based on $a^h = (a_1^h, \ldots, a_n^h)$, the players transit to the next state $s^{h+1} \sim P(\cdot \mid s^h, a^h)$. With probability $\kappa_{s^h, a^h}$, the game stops at step $h$. As $\kappa > 0$, the game will eventually stop with probability $1$. We define the random variable $H$ to be the final step before the game terminates. Additionally, we use $H_t$ to indicate the number of steps taken in episode $t$.

\paragraph{Value function and potential function}
For player $i \in [n]$, the value function $V^i(s): \Pi \rightarrow \mathbb{R}$ represents the expected cost when $s_0=s$ and the players play strategy, $(a^t)_{t\geq 0}$, $a^t \sim \pi^t, \pi^t=\left(\pi_{i}^t, \pi_{-i}^t\right)$. Specifically, it is defined as $ V_i^\pi(s):=\EE_{\pi,P, H}\left[\sum_{h=0}^{H} C_{i}^h \mid s_0=s\right] $. We slightly abuse the notation and denote $V^\pi_i(\mu)=\EE_{s \sim \mu}\left[V_i^\pi(s)\right]$. 

We define the state-action value function for player $i$ as $Q_i^\pi(s, a)$, which denotes the expected cost when started from $s_0 = s$, $a_0 = a$, $$Q_i^\pi(s, a)=\EE_{\pi,P, H}\left[\sum_{h=0}^{H}  C_i^h \mid s_{0}=s, a_{0}=a\right] \,.$$
We also denote $Q_i^\pi(\mu, a) = \EE_{s\sim \mu} \left[Q_i^\pi(s, a)\right]$. Notice that the value function can be equivalently expressed as $$V_i^\pi(s)=\sum_{a \in \mathcal{A}} \pi\left(a \mid s\right) Q_i^\pi\left(s, a\right)\,.$$ 

In the Markov potential game, there exists a global potential function $\Phi$, that captures the incentive of all players to change their strategies at any state. Specifically, at any state $s \in \mathcal{S}$, $\pi_i, \pi_i^\prime$, $i \in [n]$, the potential function satisfies, $ V_i^{\pi_i, \pi_{-i}}(s) - V_i^{\pi_i^\prime, \pi_{-i}}(s) = \Phi_s(\pi_i, \pi_{-i}) - \Phi_s(\pi_i^\prime, \pi_{-i})$.
By linearity of expectations, we also have $V_i^{\pi_i, \pi_{-i}}(\mu) - V_i^{\pi_i^\prime, \pi_{-i}}(\mu) = \Phi_\mu(\pi_i, \pi_{-i}) - \Phi_\mu(\pi_i^\prime, \pi_{-i})$,
where $\Phi_\mu(\pi_i, \pi_{-i}) = \EE_{s\sim \mu}\left[\Phi_s(\pi_i, \pi_{-i}) \right] $. 

\paragraph{Solution concepts}
Similar to the one-step potential game, the approximate Nash equilibrium is defined as follows
\begin{defn}[$\epsilon$-Nash equilibrium]
    A strategy $\pi^\ast = (\pi^\ast_1, \ldots, \pi^\ast_n)$ is called an $\epsilon$-\textbf{approximate Nash equilibirum} if for all player $i \in [n]$, it holds  $V_i(\pi_i^\ast, \pi_{-i}^\ast) \leq  V_i(\pi_i, \pi_{-i}^\ast) \,, \quad \forall \pi_i \in \Delta(\mathcal{A}_i)^\mathcal{S} $.
    When $\epsilon = 0$, we call it a \textbf{Nash equilibrium}.
\end{defn}

Such approximate Nash equilibrium can be obtained by achieving sublinear Nash regret, which is defined as follows. 
\begin{defn}[Nash regret in MPG] \label{defn:nash_regret_mpg}
Let $\{\pi^t\}$ be a sequence of stochastic policies played by the players across $T$ episodes, the Nash regret after $T$ episodes are defined as $\mathrm{Nash\text{-}regret}(T)  =   \sum^T_{t=1} \max_{i \in [n]} \left(V_i^{\pi_i^t, \pi_{-i}^t}(\mu) - \min_{\pi_i} V_i^{\pi_i, \pi_{-i}^t}(\mu)\right) $.
\end{defn}

For a selfish individual player, another important metric for playing against potentially malicious adversaries is regret, which is defined as follows. 
\begin{defn}[Regret in MPG] 
For any $i \in [n]$, and an arbitrary sequence of $\{\pi_{-i}^t\}_{t=1}^T$, the individual regret of player $i$ is defined as $ \sum^T_{t=1} \left(V_i^{\pi_i^t, \pi_{-i}^t}(\mu) - \min_{\pi_i} \sum^T_{t=1} V_i^{\pi_i, \pi_{-i}^t}(\mu)\right)$. 
\end{defn}

\subsection{Algorithm for Markov potential game}
To apply Algorithm \ref{alg} to the Markov potential game, we need to build an unbiased one-sample gradient estimate. This may be constructed from a single or more steps of actions. We use the following estimator, which can be demonstrated to be unbiased and has bounded norms:
\begin{align}
    \widehat{\nabla}_{\pi_i^t}=\sum_{h=0}^{H_t} C_i^h\sum_{h=0}^{H_t} \nabla \log \pi_i^t\left(a_i^{h} \mid s^{h}\right)\,.
\end{align}

With this, we can construct the recursive gradient estimator and perform the Frank-Wolfe update step. The overall algorithm is described in Algorithm \ref{alg:markov}.

\begin{algorithm}[ht]
\DontPrintSemicolon
\KwInput{Parameters $\eta_t$, $\rho_t$, $\mu$.}
Initialize $\pi_i^1$, $d_i^0$, for $i \in [n]$\;
\For{$t = 1, \ldots, T$}{
    $\tilde{\pi}_i^{t} = (1 - \mu)\pi_i^{t} + \mu s$, where $s$ is a uniformly random distribution on $\Delta(\mathcal{A}_i)$\;
    \For{$h = 0, \ldots, H$}{
        Play $a_i^h \sim \tilde{\pi}_i^t$ and receive rewards $C^h_i$\;
    }
    $\widehat{\nabla}_{\pi_i^t}=\sum_{h=0}^{H_t} C_i^h\sum_{h=0}^{H_t} \nabla \log \pi_i^t\left(a_i^{h} \mid s^{h}\right)$\;
    Update $d^t_i = (1-\rho_t)d^{t-1}_i + \rho_t \hat{\nabla}_{\pi_i^t}$\;
    Update $\hat{\pi}_i^{t+1} = \argmin_{\pi_i \in \Delta(\mathcal{A}_i)} \langle \pi_i, d^t_i \rangle$\;
    Update $\pi_i^{t+1} = (1-\eta_t)\pi_i^t + \eta_t \hat{\pi}_i^{t+1}$\;
}
\caption{Stochastic Frank-Wolfe with exploration for Markov potential game}\label{alg:markov}
\end{algorithm}

Equipped with Algorithm \ref{alg:markov}, the following theorem establishes the Nash regret for Markov potential game. 

\begin{restatable}[Nash regret for Markov potential game]{thm}{nashregretmpg}\label{thm:nash_mpg}
If all players run Algorithm \ref{alg:markov} with  $\eta_t =  \frac{\kappa}{n^{3/2}mt^{4/5}} $, $\rho_t = \frac{4 \mu^{1/3}}{n^{1/3}m^{2/3} t^{8/15} }$ and $\mu = \min\{\frac{1}{m^2n}, \frac{\kappa^{4/3}}{n^{7/8}m^{1/4}T^{\beta}}\}$, when $T \geq m^{7/4} n^{1/8} \kappa^{4/3}$, we have 
\begin{align*}
    & \EE\left[\mathrm{Nash\text{-}regret}(T)\right] \\
    \leq \ & O \left(\frac{ n^{3/2}mD_\infty T^{4/5}}{\kappa^{3}} + \frac{n^{5/8}m^{3/4}SD_\infty T^{4/5}}{\kappa^{25/9}}\right)  \,,
\end{align*}
where $D_\infty = \max_i \left\|\frac{d^{\pi_i^\ast, \pi_{-i}}(\mu)}{\mu}\right\|_{\infty}$.
\end{restatable}
\begin{rem}
    When $T \geq O \left( \frac{n^{17/2}m^{5} D_\infty^5}{\kappa^{15} \epsilon^5}  \frac{n^{15/8}m^{15/4} D_\infty^5 S^5}{\kappa^{25/3} \epsilon^5}\right)$, there exists $t \in \{1, \ldots, T\}$, such that $\pi^t$ is an $\epsilon$-equilibrium.
\end{rem}
\begin{rem}
    Using Markov inequality, it is immediate that Theorem \ref{thm:nash_mpg} also holds with high probability. 
\end{rem}
Our bound relies on a distribution mismatch coefficient $D_\infty = \left\|\frac{d^{\pi_i^\ast, \pi_{-i}}(\mu)}{\mu}\right\|_{\infty}$, which is an extension to the distribution mismatch coefficient ($D_{\infty}:=\max_s\left(d_{s_0}^{\pi^{\ast}}(s) / \mu(s)\right)$) \citep{agarwal2021theory}. In the single-agent case, the coefficient is large whenever an MDP is difficult to explore, which may manifest in two cases: 1) a state $s$ is important and the optimal policy needs to visit the state a lot, while the starting distribution rarely starts from this, or 2) a state $s$ is unimportant and the optimal policy rarely spends time on this state, but it is likely to start from this state. These are extended to the multi-agent setting with $d_{s_0}^{\pi^{\ast}_i, \pi_{-i}}(s)$ defined instead, which means the time spent on state $s$ by player $i$ taking the optimal policy $\pi_i^\ast$ while others play $\pi_{-i}$, to capture the exploration difficulty for individual player. 

Compared to previous results of $O (\epsilon^{-6})$ on Markov potential game, our results improve them by $O(\epsilon^{-1})$ \citep{leonardos2022global}. Assuming the value function admits a linear instruction (i.e. $Q_i^\pi(s,a) = \langle \phi(s,a), \omega_i \rangle$, where $\phi$ is a known feature mapping and $\omega_i$ is a learnable parameter), \citet{ding2022independent} established a $O(\epsilon^{-5})$ result under a regression oracle. However, the oracle can only be implemented for such a result when $\phi^\top \phi$ is invertible. With a tabular Markov potential game (where the states and actions are finite), $\phi$ may be a one-hot basis vector, and thus the assumption is not satisfied. 

\begin{figure*}[htb]
    \begin{subfigure}[b]{0.45\textwidth}
      \includegraphics[width=\textwidth]{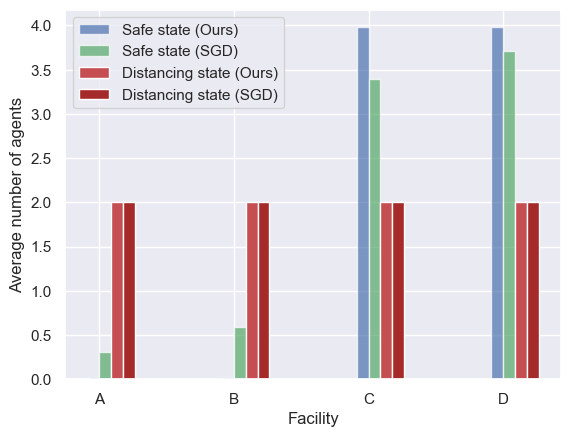}
      \caption{}\label{fig:facilities}
    \end{subfigure}
    \begin{subfigure}[b]{0.45\textwidth}
      \includegraphics[width=\textwidth]{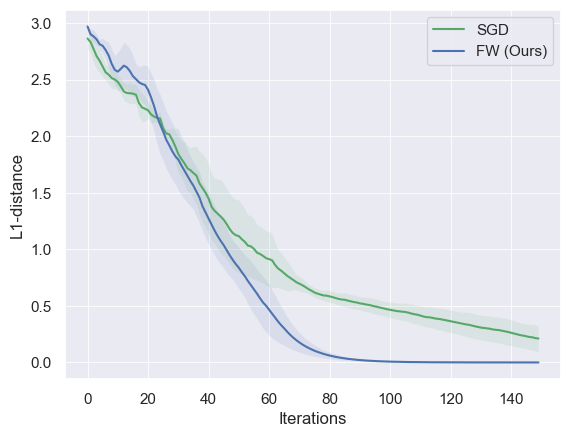}
      \caption{}\label{fig:curve}
    \end{subfigure}
    \centering
\caption{Figure \ref{fig:facilities} shows the final converged policy on each of the states. Figure \ref{fig:curve} shows the convergence of the algorithms by $L_1$ distance to the final strategy.}
\end{figure*}

We also remark that our choice of parameters $\eta_t, \rho_t, \mu$ does not depend on $D_\infty$ and hence requires no knowledge of the exploration difficulty. In contrast, the previous bounds crucially rely on the knowledge of $D_\infty$ to choose the parameters \citep{zhang2021gradient,leonardos2022global,ding2022independent}.

The following theorem establishes the regret guarantee for individual players, even when the other players are playing arbitrarily.

\begin{restatable}[Regret for the $i$-th player in Markov potential game]{thm}{individualmpg} \label{thm:individual_mpg} 
Let $\{\tilde{\pi}_{-i}^t\}_{t=1}^T$ be an arbitrary sequence of strategies for all players except for player $i$. For player $i$, run Algorithm \ref{alg:markov} with $\eta_t =  \frac{\kappa}{n^{3/2}mt^{4/5}} $, $\rho_t = \frac{4 \mu^{1/3}}{n^{1/3}m^{2/3} t^{8/15} }$ and $\mu = \min\{\frac{1}{m^2n}, \frac{\kappa^{4/3}}{n^{7/8}m^{1/4}T^{\beta}}\}$, when $T \geq m^{7/4} n^{1/8} \kappa^{4/3}$. 
Then the regret of player $i$ is at most
\begin{align*}
   & \EE\left[\sum^T_{t=1} \left(V_i^{\tilde{\pi}_i^t, \tilde{\pi}_{-i}^t}(\mu) -  V_i^{\pi_i^\ast, \tilde{\pi}_{-i}^t}(\mu)\right) \right]\\
   \leq \ & O \left(\frac{m^{4/3}n^{27/24}ST^{4/5}}{\kappa^{7/3}}\right)\,.
\end{align*}
\end{restatable}


\section{Experiments}
To validate our theoretical results, we implemented our algorithm on an example of the Markov congestion game, taken from \citet{leonardos2022global}. 

\paragraph{Experiments setup} In the example game, we have $n = 8$ players, $\mathcal{A}_i = 4$ actions for player $i$ to choose from, and $\mathcal{S} = \{\text{safe}, \text{distancing}\}$ as the set of states. In both states, each player's best strategy is to be in the same facility with other players to minimize the cost. In particular, the cost of each facility is determined by the number of players choosing it, whereas the facility preference is $D > C > B> A$ as they have increasing costs in that order. However, when more than $n/2 = 4$ players choose the same facility, the state will be transited to the distancing state. The cost structure is the same in the distancing state, but each player will be penalized more by a constant amount (by 100 times more). To return to the safe state, the players will need to learn to distribute evenly across the facilities, with no more than $n/4 = 2$ players on the same facility. The game stops at each step with a probability $0.99$. However, we limit the maximum length of the game to $H_t = 20$.
When the transition is deterministic, it is clear that the optimal strategy for the players is to choose facility $C$ and $D$ with $4$ players each in the safe state and to choose each of the facilities evenly in the distancing state. 

We note that all the results shown are the averaged results obtained from 5 runs with different random seeds; the shades in Figure \ref{fig:curve} denote $1$ standard deviation from the mean. All experiments are conducted with a $10$ core CPU and $16$ GB RAM. 

\paragraph{Implementation and results}
We implemented the projected stochastic gradient descent method (SGD) proposed by \cite{leonardos2022global} and our stochastic Frank-Wolfe with exploration. The algorithms are run with $T = 150$ iterations, and each update is performed with $10$ trajectories of fixed length $H_t = 20$. For the SGD method, we kept its learning rate as $\eta = 0.0001$ as specified in their paper. For stochastic FW with exploration, we choose $\eta = 0.1$, $\rho = \frac{0.9}{(t+1)^{3/5}}$ and $\mu = 0.001$. 

Figure \ref{fig:facilities} shows the final learned strategy of each of the algorithms. As shown in the figure, our method learns the optimal strategy in both states after the algorithm iterations. In Figure \ref{fig:curve}, we show the convergence of the algorithms by plotting the distance (in $L_1$ distance) of the strategy to the final strategy. In comparison, our algorithm converges much faster than the SGD method.

\section{Conclusion}
In this work, we studied potential games and Markov potential games with stochastic cost and bandit feedback and introduced an algorithm based on the celebrated Frank-Wolfe algorithm. Our algorithm converges to Nash equilibrium fast with a Nash regret of $O(T^{4/5})$, while achieving sublinear regret for each player. This improves over the previous Nash regret result for the Markov potential game and is the first algorithm to simultaneously achieve small Nash regret and regret. Experimental results validate the effectiveness of our algorithms. 

\section*{Acknowledgement}
We thank the reviewers and the area chair for constructive feedback. YY gratefully acknowledges funding support from NSERC and the Canada CIFAR AI chairs program. Baoxiang Wang is partially supported by the National Natural Science Foundation of China (62106213, 72150002, 72394361) and the Shenzhen Science and Technology Program (RCBS20210609104356063, JCYJ20210324120011032).

\bibliographystyle{apalike}
\bibliography{sCG}

\begin{thebibliography}{}

\bibitem[Agarwal et~al., 2021]{agarwal2021theory}
Agarwal, A., Kakade, S.~M., Lee, J.~D., and Mahajan, G. (2021).
\newblock On the theory of policy gradient methods: {O}ptimality,
  approximation, and distribution shift.
\newblock {\em The Journal of Machine Learning Research}, 22(1):4431--4506.

\bibitem[Anagnostides et~al., 2022]{anagnostides2022last}
Anagnostides, I., Panageas, I., Farina, G., and Sandholm, T. (2022).
\newblock On last-iterate convergence beyond zero-sum games.
\newblock In {\em International Conference on Machine Learning}.

\bibitem[Chen et~al., 2022]{chen2022convergence}
Chen, D., Zhang, Q., and Doan, T.~T. (2022).
\newblock Convergence and price of anarchy guarantees of the softmax policy
  gradient in markov potential games.
\newblock In {\em International Conference on Machine Learning: Workshop AI for
  Agent-Based Modelling}.

\bibitem[Chen et~al., 2009]{chen2009settling}
Chen, X., Deng, X., and Teng, S.-H. (2009).
\newblock Settling the complexity of computing two-player {N}ash equilibria.
\newblock {\em Journal of the ACM (JACM)}, 56(3):1--57.

\bibitem[Cohen et~al., 2016]{cohen2016exponentially}
Cohen, J., H{\'e}liou, A., and Mertikopoulos, P. (2016).
\newblock Exponentially fast convergence to (strict) equilibrium via hedging.
\newblock arXiv preprint arXiv:1607.08863.

\bibitem[Combes et~al., 2015]{combes2015combinatorial}
Combes, R., Talebi Mazraeh~Shahi, M.~S., Proutiere, A., et~al. (2015).
\newblock Combinatorial bandits revisited.
\newblock In {\em Advances in neural information processing systems}.

\bibitem[Cominetti et~al., 2010]{cominetti2010payoff}
Cominetti, R., Melo, E., and Sorin, S. (2010).
\newblock A payoff-based learning procedure and its application to traffic
  games.
\newblock {\em Games and Economic Behavior}, 70(1):71--83.

\bibitem[Cui et~al., 2022]{cuilearning}
Cui, Q., Xiong, Z., Fazel, M., and Du, S.~S. (2022).
\newblock Learning in congestion games with bandit feedback.
\newblock In {\em Advances in Neural Information Processing Systems}.

\bibitem[Daskalakis, 2013]{daskalakis2013complexity}
Daskalakis, C. (2013).
\newblock On the complexity of approximating a {N}ash equilibrium.
\newblock {\em ACM Transactions on Algorithms (TALG)}, 9(3):1--35.

\bibitem[Daskalakis et~al., 2020]{daskalakis2020independent}
Daskalakis, C., Foster, D.~J., and Golowich, N. (2020).
\newblock Independent policy gradient methods for competitive reinforcement
  learning.
\newblock In {\em Advances in neural information processing systems}.

\bibitem[Ding et~al., 2022]{ding2022independent}
Ding, D., Wei, C.-Y., Zhang, K., and Jovanovic, M. (2022).
\newblock Independent policy gradient for large-scale {M}arkov potential games:
  {S}harper rates, function approximation, and game-agnostic convergence.
\newblock In {\em International Conference on Machine Learning}.

\bibitem[Dong et~al., 2023]{dong2023taming}
Dong, J., Wu, J., Wang, S., Wang, B., and Chen, W. (2023).
\newblock Taming the exponential action set: Sublinear regret and fast
  convergence to nash equilibrium in online congestion games.
\newblock arXiv preprint arXiv:2306.13673.

\bibitem[Giannou et~al., 2022]{giannou2022convergence}
Giannou, A., Lotidis, K., Mertikopoulos, P., and Vlatakis-Gkaragkounis, E.-V.
  (2022).
\newblock On the convergence of policy gradient methods to {N}ash equilibria in
  general stochastic games.
\newblock {\em Advances in Neural Information Processing Systems}.

\bibitem[Giannou et~al., 2021]{giannou2021rate}
Giannou, A., Vlatakis-Gkaragkounis, E.-V., and Mertikopoulos, P. (2021).
\newblock On the rate of convergence of regularized learning in games: {F}rom
  bandits and uncertainty to optimism and beyond.
\newblock In {\em Advances in Neural Information Processing Systems}.

\bibitem[Guo et~al., 2023]{guo2023markov}
Guo, X., Li, X., Maheshwari, C., Sastry, S., and Wu, M. (2023).
\newblock {M}arkov $\alpha$-potential games: {E}quilibrium approximation and
  regret analysis.
\newblock arXiv preprint arXiv:2305.12553.

\bibitem[Heliou et~al., 2017]{heliou2017learning}
Heliou, A., Cohen, J., and Mertikopoulos, P. (2017).
\newblock Learning with bandit feedback in potential games.
\newblock In {\em Advances in Neural Information Processing Systems}.

\bibitem[Leonardos et~al., 2022]{leonardos2022global}
Leonardos, S., Overman, W., Panageas, I., and Piliouras, G. (2022).
\newblock Global convergence of multi-agent policy gradient in {M}arkov
  potential games.
\newblock In {\em International Conference on Learning Representations}.

\bibitem[Monderer and Shapley, 1996]{monderer1996potential}
Monderer, D. and Shapley, L.~S. (1996).
\newblock Potential games.
\newblock {\em Games and economic behavior}, 14(1):124--143.

\bibitem[Palaiopanos et~al., 2017]{palaiopanos2017multiplicative}
Palaiopanos, G., Panageas, I., and Piliouras, G. (2017).
\newblock Multiplicative weights update with constant step-size in congestion
  games: {C}onvergence, limit cycles and chaos.
\newblock In {\em Advances in Neural Information Processing Systems}.

\bibitem[Panageas et~al., 2023]{panageas23}
Panageas, I., Skoulakis, S., Viano, L., Wang, X., and Cevher, V. (2023).
\newblock Semi bandit dynamics in congestion games: {C}onvergence to {N}ash
  equilibrium and no-regret guarantees.
\newblock In {\em International Conference on Machine Learning}.

\bibitem[Roughgarden, 2010]{roughgarden2010algorithmic}
Roughgarden, T. (2010).
\newblock Algorithmic game theory.
\newblock {\em Communications of the ACM}, 53(7):78--86.

\bibitem[Shalev-Shwartz et~al., 2012]{shalev2012online}
Shalev-Shwartz, S. et~al. (2012).
\newblock Online learning and online convex optimization.
\newblock {\em Foundations and Trends{\textregistered} in Machine Learning},
  4(2):107--194.

\bibitem[Shapley, 1953]{shapley1953stochastic}
Shapley, L.~S. (1953).
\newblock Stochastic games.
\newblock {\em Proceedings of the National Academy of Sciences},
  39(10):1095--1100.

\bibitem[Song et~al., 2021]{song2021can}
Song, Z., Mei, S., and Bai, Y. (2021).
\newblock When can we learn general-sum markov games with a large number of
  players sample-efficiently?
\newblock In {\em International Conference on Learning Representations}.

\bibitem[Yuan et~al., 2016]{yuan2016influence}
Yuan, K., Ying, B., and Sayed, A.~H. (2016).
\newblock On the influence of momentum acceleration on online learning.
\newblock {\em The Journal of Machine Learning Research}, 17(1):6602--6667.

\bibitem[Zhang et~al., 2020]{ZhangSMHK20}
Zhang, M., Shen, Z., Mokhtari, A., Hassani, H., and Karbasi, A. (2020).
\newblock One sample stochastic {F}rank-{W}olfe.
\newblock In {\em International Conference on Artificial Intelligence and
  Statistics}.

\bibitem[Zhang et~al., 2022a]{zhang2022policy}
Zhang, R., Liu, Q., Wang, H., Xiong, C., Li, N., and Bai, Y. (2022a).
\newblock Policy optimization for {M}arkov games: {U}nified framework and
  faster convergence.
\newblock In {\em Advances in Neural Information Processing Systems}, pages
  21886--21899.

\bibitem[Zhang et~al., 2022b]{zhang2022global}
Zhang, R., Mei, J., Dai, B., Schuurmans, D., and Li, N. (2022b).
\newblock On the global convergence rates of decentralized softmax gradient
  play in {M}arkov potential games.
\newblock In {\em Advances in Neural Information Processing Systems}.

\bibitem[Zhang et~al., 2021]{zhang2021gradient}
Zhang, R., Ren, Z., and Li, N. (2021).
\newblock Gradient play in stochastic games: {S}tationary points, convergence,
  and sample complexity.
\newblock arXiv preprint arXiv:2106.00198.

\end{thebibliography}
\clearpage

\appendix
\onecolumn
\aistatstitle{Appendix for ``Convergence to Nash Equilibrium and No-regret Guarantee in (Markov) Potential Games''} 
\thispagestyle{plain}

\section{Analysis for potential game}
As each player shares the same configurations of parameters, we drop the index $i$ and rewrite the update steps as 
\begin{align*}
     \tilde{\pi}^{t} = \ &  (1 - \mu)\pi^{t} + \mu s\\
     (\hat{C}^t)_a = \ & \frac{C^t \cdot \mathbb{I} \{a = a^t\}}{(\tilde{\pi}^t)_a} \,, \quad \forall a \in \mathcal{A} \,, \quad 
    d^t =  (1-\rho_t)d^{t-1} + \rho_t \hat{C}^t \\
    \hat{\pi}^{t+1} =  \ & \argmin_{\pi \in \Delta(\mathcal{A})} \langle \pi, d^t \rangle \\
    \pi^{t+1} = \ &  (1-\eta_t)\pi^t + \eta_t \hat{\pi}^{t+1} \,.
\end{align*}


\subsection{Proof of Theorem \ref{thm:nash_regret}}
\nashregret*

\begin{proof}
We start by decomposing the regret as,
\begin{align*}
       \mathrm{Nash\text{-}regret}(T) 
        = \ & \sum^T_{t=1} \max_{i\in[n]} \left(c_i\left(\tilde{\pi}_i^t, \tilde{\pi}_{-i}^t\right) - \min_{\pi_i^\prime \in \Delta(\mathcal{A}_i)} c_i\left(\pi_i^\prime, \tilde{\pi}_{-i}^t\right)\right) \\
        \leq \ & \sum^n_{i=1} \sum^T_{t=1} \max_{\pi_i^\prime \in \Delta(\mathcal{A}_i)} \left(c_i\left(\tilde{\pi}_i^t, \tilde{\pi}_{-i}^t\right) -  c_i\left(\pi_i^\prime, \tilde{\pi}_{-i}^t\right)\right) \\
        = \ & \sum^n_{i=1} \sum^T_{t=1} \max_{\pi_i^\prime \in \Delta(\mathcal{A}_i)} \Phi\left(\tilde{\pi}_i^t, \tilde{\pi}_{-i}^t\right) -  \Phi\left(\pi_i^\prime, \tilde{\pi}_{-i}^t\right) \,,
    \end{align*}
    where the last equality is by the definition of potential function. 

    Denote $\pi_i^{t, \prime} = \argmax_{\pi_i^\prime \in \Delta(\mathcal{A}_i)} \Phi\left(\tilde{\pi}_i^t, \tilde{\pi}_{-i}^t\right) -  \Phi\left(\pi_i^\prime, \tilde{\pi}_{-i}^t\right)$, we have
    \begin{align*}
        & \max_{\pi_i^\prime} \Phi\left(\tilde{\pi}_i^t, \tilde{\pi}_{-i}^t\right) -  \Phi\left(\pi_i^\prime, \tilde{\pi}_{-i}^t\right) \\
        = \ & \left(\Phi\left(\pi_i^t, \pi_{-i}^t\right) -  \Phi\left(\pi_i^{t, \prime}, \pi_{-i}^t\right) \right) + \left(\Phi(\tilde{\pi}^t) - \Phi(\pi^t)\right) + \left(\Phi\left(\pi_i^{t, \prime}, \pi_{-i}^t\right) - \Phi\left(\pi_i^{t, \prime}, \tilde{\pi}_{-i}^t\right)\right) \\
        \leq \ & \max_{\pi_i^\prime} \Phi\left(\pi_i^t, \pi_{-i}^t\right) -  \Phi\left(\pi_i^\prime, \pi_{-i}^t\right)+ \left(\Phi(\tilde{\pi}^t) - \Phi(\pi^t)\right) + \left(\Phi\left(\pi_i^{t, \prime}, \pi_{-i}^t\right) - \Phi\left(\pi_i^{t, \prime}, \tilde{\pi}_{-i}^t\right)\right) \\
        \leq \ & \max_{\pi_i^\prime} \Phi\left(\pi_i^t, \pi_{-i}^t\right) -  \Phi\left(\pi_i^\prime, \pi_{-i}^t\right) +  \sqrt{n}\|\tilde{\pi}^t - \pi^t\|_2 + \sqrt{n} \|\pi_{-i}^t - \tilde{\pi}_{-i}^t\|_2 \\
        \leq \ & \max_{\pi_i^\prime} \Phi\left(\pi_i^t, \pi_{-i}^t\right) -  \Phi\left(\pi_i^\prime, \pi_{-i}^t\right) + \mu \sqrt{n}\|\pi^t - s\|_2 + \mu \sqrt{n}\|\pi_{-i}^t  - s_{-i}\|_2 \\
        \leq \ & \max_{\pi_i^\prime} \Phi\left(\pi_i^t, \pi_{-i}^t\right) -  \Phi\left(\pi_i^\prime, \pi_{-i}^t\right)  + 2n \mu \,,
    \end{align*}
 where the second inequality is by the Smoothness of $\Phi$ and the third inequality is by the update rule $\Tilde{\pi}^t = (1-\mu) \pi^t + \mu s$.
\bigbreak
\pagebreak
Define $G(\pi^t) = \max_\pi (\pi^t - \pi)^\top \nabla \Phi(\pi^t)$. By  the smoothness of $\Phi$, we have
    \begin{align*}
        \Phi(\pi^{t+1}) 
        \leq \ & \Phi(\pi^{t}) + (\pi^{t+1} - \pi^t)^\top \nabla \Phi(\pi^t) + \frac{L}{2} \|\pi^{t+1} - \pi^t\|_2^2 \\
        \leq \ & \Phi(\pi^{t}) + \eta_t (\hat{\pi}^{t+1} - \pi^t)^\top \nabla \Phi(\pi^t) + \frac{\eta_t^2 L}{2} \|\hat{\pi}_i^{t+1} - \pi^t\|_2^2 \\
        = \ &  \Phi(\pi^{t}) + \eta_t(\hat{\pi}^{t+1} - \pi^t)^\top  d^t + \eta_t(\hat{\pi}^{t+1} - \pi^t)^\top  \left(\nabla \Phi(\pi^t)  - d^t\right) + \frac{\eta_t^2 L}{2} \|\hat{\pi}_i^{t+1} - \pi^t\|_2^2 \\
        \leq \ & \Phi(\pi^{t}) + \eta_t(\hat{\pi}^{t+1} - \pi^t)^\top  d^t + \eta_t \|\hat{\pi}^{t+1} - \pi^t\|_2 \left\|\nabla \Phi(\pi^t)  - d^t \right\|_2 + \frac{\eta_t^2 L}{2} \|\hat{\pi}_i^{t+1} - \pi^t\|_2^2 \\
        \leq \ & \Phi(\pi^{t}) + \eta_t(\hat{\pi}^{t+1} - \pi^t)^\top d^t + \eta_t \sqrt{n}  \left\|\nabla \Phi(\pi^t)  - d^t\right\|_2 + \frac{\eta_t^2 n L}{2}\,.
    \end{align*}
    where the second inequality is by update rule $\hat{\pi}^{t+1} = \argmin_{\pi \in \Delta(\mathcal{A})} \langle \pi, d^t \rangle$, the third inequality is by Cauchy-Schwarz inequality, and the last inequality is by Cauchy-Schwarz and that $\|\pi\|_2 \leq \sqrt{n}$ for any $\pi \in \Delta(\mathcal{A})$.

    Let $\bar{\pi}^{t+1} = \argmin_{\pi} \langle \pi, \nabla \Phi(\pi^t)\rangle$.
    Then 
    \begin{align*}
        & \Phi(\pi^{t+1}) \\
        \leq \ &  \Phi(\pi^{t}) + \eta_t(\hat{\pi}^{t+1} - \pi^t)^\top d^t + \eta_t  \sqrt{n}  \left\|\nabla \Phi(\pi^t)  - d^t\right\|_2 + \frac{\eta_t^2 n L}{2} \\
        \leq \ & \Phi(\pi^{t}) + \eta_t(\bar{\pi}^{t+1} - \pi^t)^\top d^t + \eta_t  \sqrt{n}  \left\|\nabla \Phi(\pi^t)  - d^t\right\|_2 + \frac{\eta_t^2 n L}{2} \\
        = \ & \Phi(\pi^{t}) + \eta_t(\bar{\pi}^{t+1} - \pi^t)^\top \nabla \Phi(\pi^t) + \eta_t(\Tilde{x}_i^{t+1} - \pi^t)^\top \left(d^t - \nabla \Phi(\pi^t) \right)+ \eta_t  \sqrt{n}  \left\|\nabla \Phi(\pi^t)  - d^t\right\|_2 + \frac{\eta_t^2 n L}{2}  \\
        \leq \ &  \Phi(\pi^{t}) - \eta_t G(\pi^t) + 2\eta_t  \sqrt{n}  \left\|\nabla \Phi(\pi^t)  - d^t\right\|_2 + \frac{\eta_t^2 n L}{2}  \,,
    \end{align*}
    where the second inequality is by the update rule, and the third inequality is by the definition of $G(\pi^t)$ and Cauchy-Schwarz. 

    Rearrange the terms gives 
    \begin{align*}
        G(\pi^t) \leq \ & \frac{\Phi(\pi^{t}) - \Phi(\pi^{t+1})}{\eta_t} +2  \sqrt{n} \left\|\nabla \Phi(\pi^t)  - d^t\right\|_2 + \frac{\eta_t n L}{2} \,.
    \end{align*}

    Summing over $T$ gives
    \begin{align*}
        \sum^{T}_{t=1} G(\pi^t) 
        \leq \ & \sum^{T}_{t=1}\frac{\Phi(\pi^{t}) -  \Phi(\pi^{t+1})}{\eta_t} + 2 \sqrt{n}  \sum^{T}_{t=1} \left\|\nabla \Phi(\pi^t)  - d^t\right\|_2 + \frac{n L}{2}\sum^{T}_{t=1}\eta_t  \\
        = \ & \sum^{T}_{t=1} \left(\left(\frac{ \Phi(\pi^{t})}{\eta_t} - \frac{\Phi(\pi^{t+1})}{\eta_{t+1}}\right) + \left(\frac{1}{\eta_{t+1}} - \frac{1}{\eta_{t}}\right)\Phi(\pi^{t+1})\right) + 2 \sqrt{n} \sum^{T}_{t=1} \left\|\nabla \Phi(\pi^t)  - d^t\right\|_2 + \frac{n L}{2}\sum^{T}_{t=1}\eta_t \\
        \leq \ & \left(\frac{\Phi(\pi^{1})}{\eta_1} - \frac{\Phi(\pi^{T+1})}{\eta_{T+1}}\right) +  \frac{n}{\eta_{T+1}} + 2 \sqrt{n} \sum^{T}_{t=1} \left\|\nabla \Phi(\pi^t)  - d^t\right\|_2 + \frac{n L}{2}\sum^{T}_{t=1}\eta_t \\
        \leq \ & \frac{2n}{\eta_{T+1}} + 2 \sqrt{n}  \sum^{T}_{t=1} \left\|\nabla \Phi(\pi^t)  - d^t\right\|_2 +  \frac{n L}{2}\sum^{T}_{t=1}\eta_t\,,
    \end{align*}
    where the second inequality is because $\sum^T_{t=1}\frac{1}{\eta_{t+1}} - \frac{1}{\eta} \leq \frac{1}{\eta_{T+1}}$, and the last inequality is by $\Phi(\pi) \in [0, n]$.

    By Lemma \ref{lem:properties}, we have
    \begin{align*}
        \max_{\pi_i^\prime \in \Delta(\mathcal{A}_i)} \Phi\left(\pi_i^t, \pi_{-i}^t\right) -  \Phi\left(\pi_i^\prime, \pi_{-i}^t\right) 
        = \ & \max_{\pi_i^\prime \in \Delta(\mathcal{A}_i)}\left(\pi^t_i - \pi_i^\prime\right)^\top \nabla_{\pi_i}\Phi\left(\pi_i^t, \pi_{-i}^t\right) \\
        \leq \ &  \max_{\pi^\prime}\left(\pi^t - \pi^\prime\right)^\top \nabla\Phi\left(\pi^t\right) \\
        = \ & G(\pi^t) \,.
    \end{align*}

    Thus, for a fixed $i \in [m]$, we have
    \begin{align*}
        \sum^{T}_{t=1}\max_{\pi_i^\prime \in \Delta(\mathcal{A}_i)} \Phi\left(\pi_i^t, \pi_{-i}^t\right) -  \Phi\left(\pi_i^\prime, \pi_{-i}^t\right) 
        \leq \ & \frac{2n}{\eta_{T+1}} + 2 \sqrt{n}  \sum^{T}_{t=1} \left\|\nabla \Phi(\tilde{\pi}^t)  - d^t\right\| + \frac{n L}{2}\sum^{T}_{t=1}\eta_t \,.
    \end{align*}

    Take $\eta_t = \frac{1}{\sqrt{n}t^{3\alpha/2}}$ and $\rho_t = \frac{4\mu^{1/3}n^{1/3} m^{1/3}}{ t^\alpha }$, by Lemma \ref{lem:est_diff}, we have
    \begin{align*}
        \sqrt{n}\sum^{T}_{t=1}\EE\left[ \left\|\nabla \Phi(\pi^t)  - d^t\right\|_2\right]
        \leq \ & O \left(\mu nL T +  \frac{m^{4/3}n^{7/6}T^{1-\alpha/2}}{\mu^{1/3}} \right)\,,
    \end{align*}
    where $C$ is an absolute constant. 

    Putting everything together, and take $\mu = \min\left\{\frac{1}{mn}, \frac{m}{n^{1/8}T^\beta}\right\}$ we have
    \begin{align*}
        & \EE\left[\sum^T_{t=1} \max_{i \in [n]} \left(c_i\left(\tilde{\pi}_i^t, \tilde{\pi}_{-i}^t\right) - \min_{\pi_i^\prime \in \Delta(\mathcal{A}_i)} c_i\left(\pi_i^\prime, \tilde{\pi}_{-i}^t\right)\right) \right]\\
        = \ &  O \left(\mu n^2L T +  \frac{m^{4/3}n^{11/6}T^{1-\alpha/2}}{\mu^{1/3}}+ n^2 T^{3\alpha / 2} + n^2 L T^{1-3\alpha/2} \right)\\
        = \ & O \left(m n^{15/8}L T^{1-\beta} +  m n^{15/8}T^{1-\alpha/2+\beta/3}+ n^{2.5} T^{3\alpha / 2} + n^{2.5} L T^{1-3\alpha/2} \right) \,.
    \end{align*}

    Take $\beta = 1/5$, $\alpha = 8/15$, we have
    \begin{align*}
        \EE\left[\sum^T_{t=1} \max_{i \in [n]} \left(c_i\left(\tilde{\pi}_i^t, \tilde{\pi}_{-i}^t\right) - \min_{\pi_i^\prime \in \Delta(\mathcal{A}_i)} c_i\left(\pi_i^\prime, \tilde{\pi}_{-i}^t\right)\right) \right]
        = O \left(\left(m n^{15/8}L + n^{2.5}\right) T^{4/5}   \right)  \,.
    \end{align*}
    
\end{proof}

\newpage
\subsection{Proof of Theorem \ref{thm:individual}}
\individualregret*

\begin{proof}
Denote $\pi^\ast = \argmin_\pi \sum^T_{t=1} c_i(\pi, \tilde{\pi}_{-i}^t)$, by the definition of the potential function, we have
\begin{align*}
    \sum^T_{t=1} c_i(\tilde{\pi}_i^t, \tilde{\pi}_{-i}^t) -  c_i(\pi_i^\ast, \tilde{\pi}_{-i}^t)
    = \ & \sum^T_{t=1} \Phi(\tilde{\pi}_i^t, \tilde{\pi}_{-i}^t) -  \Phi(\pi_i^\ast, \tilde{\pi}_{-i}^t) \\
    = \ & \sum^T_{t=1} \left(\Phi(\tilde{\pi}_i^t, \tilde{\pi}_{-i}^t) -  \Phi(\pi_i^t, \tilde{\pi}_{-i}^t) \right) + \left(\Phi(\pi_i^t, \tilde{\pi}_{-i}^t) -  \Phi(\pi_i^\ast, \tilde{\pi}_{-i}^t)\right) \\
    \leq \ & \sum^T_{t=1} \|\tilde{\pi}_i^t - \pi_i^t\|_2 + \left(\Phi(\pi_i^t, \tilde{\pi}_{-i}^t) -  \Phi(\pi_i^\ast, \tilde{\pi}_{-i}^t)\right) \\
    \leq \ & \mu T + \left(\Phi(\pi_i^t, \tilde{\pi}_{-i}^t) -  \Phi(\pi_i^\ast, \tilde{\pi}_{-i}^t)\right) \,,
\end{align*}
where the last inequality is by the update rule $\tilde{\pi}^{t}_i = (1 - \mu)\pi^{t}_i + \mu s_i$ and Lemma \ref{lem:properties}. 

By the smoothness of $\Phi$, we have
\begin{align*}
    & \Phi(\pi_i^{t+1}, \tilde{\pi}_{-i}^t) -  \Phi(\pi_i^\ast, \tilde{\pi}_{-i}^t) \\
    = \ & \Phi(\pi_i^{t} + \eta_t (\hat{\pi}_i^{t+1} - \pi_i^{t}), \tilde{\pi}_{-i}^t) -  \Phi(\pi_i^\ast, \tilde{\pi}_{-i}^t)\\
    \leq \ & \left(\Phi(\pi_i^{t}, \tilde{\pi}_{-i}^t) -  \Phi(\pi_i^\ast, \tilde{\pi}_{-i}^t)\right)+ \eta_t \langle\hat{\pi}_i^{t+1} - \pi_i^t , \nabla_{\pi_i} \Phi(\pi_i^{t}, \tilde{\pi}_{-i}^t)\rangle + \frac{\eta_t^2 L}{2} \|\hat{\pi}_i^{t+1} - \pi_i^t \|^2\\
    \leq \ & \left(\Phi(\pi_i^{t}, \tilde{\pi}_{-i}^t) -  \Phi(\pi_i^\ast, \tilde{\pi}_{-i}^t)\right)+ \eta_t \langle\hat{\pi}_i^{t+1} - \pi_i^t , \nabla_{\pi_i} \Phi(\pi_i^{t}, \tilde{\pi}_{-i}^t)\rangle + \frac{\eta_t^2 L}{2}\,,
\end{align*}
where the last inequality is because $\pi$ is the concatenation of $n$ probability vectors. 

For the second term, notice that
\begin{align*}
    \langle\hat{\pi}_i^{t+1} - \pi_i^t  , \nabla_{\pi_i} \Phi(\pi_i^{t}, \tilde{\pi}_{-i}^t)\rangle
    = \ & \langle\hat{\pi}_i^{t+1} - \pi_i^t  , d^t_i\rangle + \langle\hat{\pi}_i^{t+1} - \pi_i^t  , \nabla_{\pi_i} \Phi(\pi_i^{t}, \tilde{\pi}_{-i}^t) - d^t_i\rangle \\
    \leq \ & \langle \pi_i^\ast - \pi_i^t , d^t\rangle +  \| \nabla_{\pi_i} \Phi(\pi_i^{t}, \tilde{\pi}_{-i}^t) - d_i^t\|_2 \\
    \leq \ &  \langle \pi_i^\ast - \pi_i^t ,\nabla_{\pi_i} \Phi(\pi_i^{t}, \tilde{\pi}_{-i}^t) \rangle + \langle \pi_i^\ast - \pi_i^t , d^t - \nabla_{\pi_i} \Phi(\pi_i^{t}, \tilde{\pi}_{-i}^t)\rangle  + \| \nabla_{\pi_i} \Phi(\pi_i^{t}, \tilde{\pi}_{-i}^t) - d_i^t\| \\
    \leq \ & \left(\Phi(\pi_i^\ast, \tilde{\pi}_{-i}^t) - \Phi(\pi_i^{t}, \tilde{\pi}_{-i}^t)\right) + 2 \| \nabla_{\pi_i} \Phi(\pi_i^{t}, \tilde{\pi}_{-i}^t) - d_i^t\|_2 \,,
\end{align*}
where the first inequality is by the update rule, $\hat{\pi}_i^{t+1} = \argmin \langle \pi, d^t_i \rangle$, and the last inequality is because of the definition of $\Phi$

Therefore, we have
\begin{align*}
     \Phi(\pi_i^{t+1}, \tilde{\pi}_{-i}^t) -  \Phi(\pi_i^\ast, \tilde{\pi}_{-i}^t)
    \leq \ & (1 - \eta_t) \left(  \Phi(\pi_i^{t}, \tilde{\pi}_{-i}^t) -  \Phi(\pi_i^\ast, \tilde{\pi}_{-i}^t)\right) + 2 \eta_t\| \nabla_{\pi_i} \Phi(\pi_i^{t}, \tilde{\pi}_{-i}^t) - d_i^t\|_2 + \frac{\eta_t^2 L}{2} \,.
\end{align*}
By Lemma \ref{lem:est_diff}, take $\eta_t = \frac{2}{\sqrt{n}t^{3\alpha/2}}$, $\rho_t = \frac{4\mu^{1/3}}{n^{2/3} m^{2/3} t^{8/15}}$, and $\mu = \min\left\{\frac{1}{mn},\frac{m}{n^{1/8}T^{1/5}}\right\}$, we have
\begin{align*}
    & \EE \left[\Phi(\pi_i^{t+1}, \tilde{\pi}_{-i}^t) -  \Phi(\pi_i^\ast, \tilde{\pi}_{-i}^t) \right] \\ 
    \leq \ & \left(1 - \frac{2}{\sqrt{n}t^{3\alpha/2}}\right) \EE\left[\Phi(\pi_i^{t}, \tilde{\pi}_{-i}^t) -  \Phi(\pi_i^\ast, \tilde{\pi}_{-i}^t)\right] + O \left(\frac{\eta_t^2 L}{2} + \eta_t\mu \sqrt{n}L +  \frac{\eta_t m^{4/3}n^{4/3}}{\mu^{1/3}(t+1)^{\alpha/2}}\right)   
\end{align*}

Denote $a_{t+1} = \EE \left[\Phi(\pi_i^{t+1}, \tilde{\pi}_{-i}^t) -  \Phi(\pi_i^\ast, \tilde{\pi}_{-i}^t) \right] $, we have
\begin{align*}
    a_{t+1} \leq \left(1 - \frac{2}{\sqrt{n}t^{3\alpha/2}}\right)a_t + O \left(\frac{L}{t^{3\alpha}} + \frac{mn^{3/8}L}{t^{3\alpha/2 + 1/5}} +  \frac{mn}{t^{2\alpha-1/15}}\right) 
\end{align*}

Take $\alpha = 8/15$, $\beta = 1/5$, we have
\begin{align*}
    a_{t+1} 
    \leq \ & \left(1 - \frac{2}{\sqrt{n}t^{4/5}}\right)a_t +  O \left(\frac{mnL}{t}\right) \,.
\end{align*}

We now show that $a_t \leq D t^{-1/5}$ by induction, where $D = \max\{a_1, mnL\}\}$. 

When $t = 1$, we have $a_1 \leq D /1 $, which is true by the definition of D. Assume $a_t \leq D t^{-1/5}$, for $t \geq 2$. Then for $t + 1$, we have
\begin{align*}
    a_{t+1} 
    \leq \ & \left(1 - \frac{2}{\sqrt{n}t^{4/5}}\right)\frac{D}{t^{1/5}} +   O \left(\frac{mnL}{t}\right) \\
    \leq \ &  \left(1 - \frac{2}{t^{4/5}}\right)\frac{mnL}{t^{1/5}} + O \left(\frac{mnL}{t}\right) \\
    = \ & O \left(\frac{mnL}{t^{1/5}} - \frac{mnL}{t} \right)
    = O \left(\frac{mnL}{t^{1/5}} \left(\frac{t^{4/5} - 1}{t^{4/5}}\right) \right)= O \left(\frac{mnL}{t^{1/5}} \left(\frac{1}{t^{4/5} +1}\right)\right)\\
    = \ & O \left(\frac{mnL}{t^{1/5} + t}\right) \leq O \left(\frac{mnL}{t^{1/5} + 1}\right) \leq O \left(\frac{mnL}{(t+1)^{1/5} } \right) \leq O \left(\frac{D}{(t+1)^{1/5} } \right)\,.
\end{align*}

Putting everything together, for $T \geq m n^{7/8}$, we have
\begin{align*}
    \EE\left[\sum^T_{t=1} c_i(\tilde{\pi}_i^t, \tilde{\pi}_{-i}^t) -  c_i(\pi_i^\ast, \tilde{\pi}_{-i}^t)\right]
    \leq \ & O \left(mT^{4/5} + mnL\sum^T_{t=1} t^{-1/5} \right)= O \left(mnL T^{4/5}\right) \,.
\end{align*}

\end{proof}

\newpage
\subsection{Proof of Lemma \ref{lem:est_diff}}
\estdiff*


\begin{proof}
We start by decomposing the term as
    \begin{align*}
        \left\|\nabla\Phi(\pi^t) - d^t\right\|_2
        \leq \ & \left\|\nabla\Phi(\pi^t) - \nabla\Phi(\Tilde{\pi}^t)\right\|_2 + \left\|\nabla\Phi(\Tilde{\pi}^t) - d^t\right\|_2 \\
        \leq \ & L \left\|\pi^t - \tilde{\pi}^t\right\|_2 + \left\|\nabla\Phi(\Tilde{\pi}^t) - d^t\right\|_2 \\
        = \ & \mu L \|\pi^t - s\|_2 + \left\|\nabla\Phi(\Tilde{\pi}^t) - d^t\right\|_2 \\
        \leq \ & \mu L\sqrt{n} + \left\|\nabla\Phi(\Tilde{\pi}^t) - d^t\right\|_2\,,
    \end{align*}
    where the second inequality is by the smoothness of $\Phi$, the equality is by the update rule $\tilde{\pi}^{t} = (1 - \mu)\pi^{t} + \mu s$ and the last inequality is due to $\pi$ being a $n$-dimensional vector where each entry is bounded between $0$ and $1$. 

    For the second term, by the update rule of the algorithm ($d^t =  (1-\rho_t)d^{t-1} + \rho_t \hat{C}^t$), we have
    \begin{align*}
        & \left\|\nabla\Phi(\Tilde{\pi}^t) - d^t\right\|^2_2 \\
        = \ & \left\|\nabla\Phi(\Tilde{\pi}^t) - (1-\rho_t)d^{t-1} - \rho_t \hat{C}^t\right\|^2_2 \\
        = \ & \left\|\rho_t\left(\nabla\Phi(\Tilde{\pi}^t) - \hat{C}^t\right) + (1-\rho_t) \left( \nabla\Phi(\Tilde{\pi}^{t-1}) - d^{t-1}\right) + (1-\rho_t) \left(\nabla\Phi(\Tilde{\pi}^{t}) - \nabla\Phi(\Tilde{\pi}^{t-1}\right)\right\|_2^2 \\
        = \ & \rho_t^2 \left\|\nabla\Phi(\Tilde{\pi}^{t}) - \hat{C}^t\right\|_2^2 + (1-\rho_t)^2 \left\|\nabla\Phi(\Tilde{\pi}^{t-1}) - d^{t-1}\right\|_2^2 + (1-\rho_t)^2\left\| \nabla\Phi(\Tilde{\pi}^{t}) - \nabla\Phi(\Tilde{\pi}^{t-1})\right\|_2^2 \\
        & \ + 2 \rho_t(1-\rho_t) \left\langle \nabla\Phi(\Tilde{\pi}^{t-1}) - d^{t-1} ,\nabla\Phi(\Tilde{\pi}^{t}) - \hat{C}^t\right\rangle  + 2\rho_t(1 - \rho_t) \left\langle \nabla\Phi(\Tilde{\pi}^{t}) - \hat{C}^t , \nabla\Phi(\Tilde{\pi}^{t}) - \nabla\Phi(\Tilde{\pi}^{t-1}) \right\rangle\\
        & \ + 2 (1-\rho_t)^2 \left\langle \nabla\Phi(\Tilde{\pi}^{t-1}) - d^{t-1},  \nabla\Phi(\Tilde{\pi}^{t}) - \nabla\Phi(\Tilde{\pi}^{t-1})\right\rangle \,.
    \end{align*}

Let $\EE_{t}[\cdot]$ denote the conditional expectation given the history up to time $t$. 
Taking expectations on both sides and recalling $\hat{C}^t$ is an unbiased estimate of the gradient, we have
\begin{align*}
    & \EE\left[\left\||\nabla\Phi(\Tilde{\pi}^{t}) - d^t \right\|_2^2\right] \\
    = \ &  \rho_t^2 \EE\left[\left\|\nabla\Phi(\Tilde{\pi}^{t}) - \hat{C}^t\right\|_2^2 \right] + (1-\rho_t)^2 \EE\left[\left\|\nabla\Phi(\Tilde{\pi}^{t-1}) - d^{t-1}\right\|_2^2 \right]+ (1-\rho_t)^2\EE \left[\left\| \nabla\Phi(\Tilde{\pi}^{t}) - \nabla\Phi(\Tilde{\pi}^{t-1})\right\|_2^2 \right] \\
        & \ + 2 \rho_t(1-\rho_t) \EE\left[ \EE_t \left[\left\langle \nabla\Phi(\Tilde{\pi}^{t-1}) - d^{t-1} ,\nabla\Phi(\Tilde{\pi}^{t}) - \hat{C}^t\right\rangle \right] \right]\\
        & \ + 2\rho_t(1 - \rho_t)  \EE\left[\EE_t \left[\left\langle \nabla\Phi(\Tilde{\pi}^{t}) - \hat{C}^t , \nabla\Phi(\Tilde{\pi}^{t}) - \nabla\Phi(\Tilde{\pi}^{t-1}) \right\rangle \right]\right] \\
        & \ + 2 (1-\rho_t)^2  \EE\left[\left\langle \nabla\Phi(\Tilde{\pi}^{t-1}) - d^{t-1},  \nabla\Phi(\Tilde{\pi}^{t}) - \nabla\Phi(\Tilde{\pi}^{t-1})\right\rangle \right]\\
    = \ & (1-\rho_t)^2 \EE \left[\left\|\nabla\Phi(\Tilde{\pi}^{t-1}) - d^{t-1}\right\|_2^2 \right] + \rho_t^2\EE\left[\left\|\nabla\Phi(\Tilde{\pi}^{t}) - \hat{C}^t\right\|^2_2\right]  + (1-\rho_t)^2 \EE\left[\left\|\nabla\Phi(\Tilde{\pi}^{t}) - \nabla\Phi(\Tilde{\pi}^{t-1})\right\|_2^2 \right] \\
    & \ + 2 (1-\rho_t)^2 \EE \left[\left\langle \nabla\Phi(\Tilde{\pi}^{t-1}) - d^{t-1},  \nabla\Phi(\Tilde{\pi}^{t}) - \nabla\Phi(\Tilde{\pi}^{t-1})\right\rangle \right] \,.
\end{align*}

Using $2\langle a, b\rangle \leq \beta_t \|a\|_2^2 + \|b\|_2^2 / \beta_t$, for $\beta_t > 0$, we have
\begin{align*}
    \EE\left[\left\|\nabla\Phi(\Tilde{\pi}^{t}) - d^t \right\|_2^2\right]  
    \leq \ & (1-\rho_t)^2 \EE \left[\left\|\nabla\Phi(\Tilde{\pi}^{t-1}) - d^{t-1}\right\|_2^2 \right] + \rho_t^2\EE\left[\left\|\nabla\Phi(\Tilde{\pi}^{t}) - \hat{C}^t\right\|_2^2\right] \\
    & \ + (1-\rho_t)^2 \EE\left[\left\|\nabla\Phi(\Tilde{\pi}^{t}) - \nabla\Phi(\Tilde{\pi}^{t-1})\right\|_2^2 \right]  + \beta_t(1-\rho_t)^2 \EE \left[\| \nabla\Phi(\Tilde{\pi}^{t-1}) - d^{t-1} \|_2^2\right] \\
    & \ + (1-\rho_t)^2/\beta_t\EE \left[\| \nabla\Phi(\Tilde{\pi}^{t}) - \nabla\Phi(\Tilde{\pi}^{t-1})\|_2^2\right]\\
    \leq \ & (1-\rho_t)(1+\beta_t) \EE \left[\left\|\nabla\Phi(\Tilde{\pi}^{t-1}) - d^{t-1}\right\|_2^2 \right] + \rho_t^2\EE\left[\left\|\nabla\Phi(\Tilde{\pi}^{t}) - \hat{C}^t\right\|_2^2\right] \\
    & \ + (1-\rho_t)(1 + 1/\beta_t) \EE\left[\left\|\nabla\Phi(\Tilde{\pi}^{t}) - \nabla\Phi(\Tilde{\pi}^{t-1})\right\|_2^2 \right] \,.
\end{align*}

Set $\beta_t = \rho_t / 2$, we have
\begin{align*}
    \EE\left[\left\|\nabla\Phi(\Tilde{\pi}^{t}) - d^t \right\|_2^2\right] 
    \leq \ & (1-\rho_t/2)\EE \left[\left\|\nabla\Phi(\Tilde{\pi}^{t-1}) - d^{t-1}\right\|_2^2 \right] + \rho_t^2\EE\left[\left\|\nabla\Phi(\Tilde{\pi}^{t}) - \hat{C}^t\right\|_2^2\right] \\
    & \ + (1 + 2/\rho_t) \EE\left[\left\|\nabla\Phi(\Tilde{\pi}^{t}) - \nabla\Phi(\Tilde{\pi}^{t-1}) \right\|_2^2 \right] \\
    \leq \ & (1-\rho_t/2) \EE \left[\left\|\nabla\Phi(\Tilde{\pi}^{t-1}) - d^{t-1}\right\|_2^2 \right] + \rho_t^2\left(m^2 n^2 + \frac{n^2m}{\mu}\right) + (1 + 2/\rho_t) L^2\EE\left[\left\| \Tilde{\pi}^t - \Tilde{\pi}^{t-1}\right\|_2^2 \right] \\
    \leq \ & (1-\rho_t/2)\EE \left[\left\|\nabla\Phi(\Tilde{\pi}^{t-1}) - d^{t-1}\right\|_2^2 \right] + \frac{2\rho_t^2 m^2 n^2}{\mu } + (1 + 2/\rho_t)L^2 \EE\left[\left\|\pi^{t} - \pi^{t-1}\right\|_2^2 \right]  \,,
\end{align*}
where the second inequality is by the smoothness of $\Phi$,  \Cref{lem:properties} and  \Cref{lem:est_bound}.

By the update rule $\pi^{t+1} = (1-\eta_t)\pi^t + \eta_t \hat{\pi}^{t+1} $, we have $\left\|\pi_i^{t} - \pi_i^{t-1} \right\|_2^2  \leq  \eta_t^2 \left\|  \hat{\pi}_i^t - \pi_i^{t-1} \right\|_2^2 \leq \eta_t^2 n $.

Therefore, we have
\begin{align*}
    \EE\left[\left\|\nabla\Phi(\Tilde{\pi}^{t}) - d^t \right\|_2^2\right] 
    \leq \ & (1-\rho_t/2)\EE \left[\left\|\nabla\Phi(\Tilde{\pi}^{t-1}) - d^{t-1}\right\|_2^2 \right] + \frac{2\rho_t^2 m^2 n^2}{\mu } + \frac{3\eta_t^2 n}{\rho_t }\,.
\end{align*}

Define $a_t =  \EE\left[\left\|\nabla\Phi(\Tilde{\pi}^{t}) - d^t  \right\|_2^2 \right] $ and $\eta_t =  \frac{1}{\sqrt{n}t^{3\alpha/2}}$, we have
\begin{align*}
    a_t 
    \leq \ & \left(1- \frac{\rho_t}{2}\right) a_{t-1} + \frac{2\rho_t^2 m^2 n^2}{\mu } + \frac{3\eta_t^2 n}{\rho_t }\\
    \leq \ & \left(1- \frac{\rho_t}{2}\right) a_{t-1} + \frac{2\rho_t^2 m^2 n^2}{\mu } + \frac{3 }{\rho_t t^{3\alpha}} \,.
\end{align*}
Take $\rho_t = \frac{4\mu^{1/3}n^{1/3} m^{1/3}}{ t^\alpha }$. We want to show $a_t = \frac{C}{(t + 1)^{\alpha} } $, $C = \max\left\{a_1 2^{\alpha} , \frac{35m^{8/3}n^{8/3}}{\mu^{2/3}}\right\}$. We show this by induction. 

The base case $t = 1$ clearly holds due to our choice of $C_1$. 
When $\mu \geq 1/ mn$,  assume the $a_{\ell-1} \leq \frac{C}{\ell^{\alpha}}$, $2 \leq \ell \leq T$, we have
\begin{align*}
    a_\ell
      \leq \ & \left(1 - \frac{2 \mu^{1/3}n^{1/3}m^{1/3}}{\ell^\alpha}\right) \frac{C}{\ell^\alpha} + \frac{32 m^{8/3} n^{8/3}}{\mu^{1/3} \ell^{2\alpha}} + \frac{3 n^{8/3}m^{8/3}}{4  \mu^{1/3} \ell^{2\alpha}}\\
      \leq \ &  \left(1 - \frac{2}{\ell^\alpha}\right) \frac{C}{\ell^\alpha} + \frac{C}{\ell^{2\alpha}} 
     \leq C \left(\frac{\ell^\alpha - 1}{\ell^{2\alpha}}\right) \\
     = \ & \left(\frac{C}{\ell^{\alpha}+ 1}\right)  \leq  \left(\frac{C}{(\ell+1)^{\alpha}}\right)  \,.
\end{align*}

Taking square root on both sides yield,  
\begin{align*}
     \EE \left[\|\nabla \Phi(\tilde{\pi}^t) - d^t \|_2 \right] \leq \frac{\max\{C_1, \sqrt{35}m^{4/3}n^{4/3}\}}{\mu^{1/3}(t+1)^{\alpha/2}}\,,
\end{align*}
where $C_1$ is a constant.
Combining everything yields the claimed result. 
\end{proof}

\newpage
\subsection{Auxiliary Lemmas}
\begin{lem}\label{lem:est_bound}
    Let $\EE_t[\cdot]$ denote the conditional expectation given the history up to time $t$, we have
    \begin{itemize}
        \item $ \EE_t[(\hat{C}^t_i)_a] =  (\nabla_{\pi_i} \Phi(\tilde{\pi}))_a$.
        \item $\EE_t[\|\hat{C}^t_i\|^2] = \frac{m}{\mu}$.
    \end{itemize}
\end{lem}
\begin{proof}
Let $\EE_t[\cdot]$ denote the conditional expectation given the history up to time $t$. Then we have 
    \begin{align*}
        \EE_t[(\hat{C}^t_i)_a]
        =  \EE_t \left[\frac{C^t_i \cdot \mathbb{I} \{a = a_i^t\}}{(\tilde{\pi}_i^t)_a}\right]
        = \frac{c_i(a, \tilde{\pi}_{-i})}{(\tilde{\pi}_i^t)_a} \EE_t [\mathbb{I} \{a = a_i^t\}]
        = \frac{c_i(a, \tilde{\pi}_{-i})}{(\tilde{\pi}_i^t)_a} (\pi_i^t)_a
        = c_i(a, \tilde{\pi}_{-i}) = (\nabla_{\pi_i} \Phi(\tilde{\pi}))_a \,.
    \end{align*}
    We also have
    \begin{align*}
        \EE_t[\|\hat{C}^t_i\|^2]
        \leq \ &  \sum_{a \in \mathcal{A}_i}\EE_t\left[\left(\frac{C^t_i \cdot \mathbb{I} \{a = a_i^t\}}{(\tilde{\pi}_i^t)_a}\right)^2\right]
        = \  \sum_{a \in \mathcal{A}_i} c_i(a, \tilde{\pi}_{-i})^2 \EE_t\left[\frac{1}{(\tilde{\pi}_i^t)_a^2}\right]
        = \ \sum_{a \in \mathcal{A}_i} c_i(a, \tilde{\pi}_{-i})^2 \frac{(\tilde{\pi}_i^t)_a}{(\tilde{\pi}_i^t)_a^2}
        \leq \ \frac{m}{\mu} \,,
    \end{align*}
    where the last inequality is by our assumption $c_i(a, \tilde{\pi}_{-i}) \in [0,1]$.
\end{proof}

\begin{lem}[Properties of the potential function] \label{lem:properties}
For any $\pi, \pi^\prime$, we have
    \begin{enumerate}
        \item $ \|\Phi(\pi) - \Phi(\pi^\prime)\|_2 \leq \sqrt{n}\| \pi - \pi^\prime\|_2 $.
        \item $\|\nabla_{\pi_i} \Phi(\pi)\|_2 \leq m$.
        \item $\Phi(\pi_i, \pi_{-i}) - \Phi(\pi^\prime_i, \pi_{-i})
        = (\pi_i - \pi_i^\prime) \nabla_{\pi_i} \Phi(\pi_i, \pi_{-i})$.
    \end{enumerate}
\end{lem}
\begin{proof}
The first one is true as $\|\nabla_\pi \Phi(\pi)\|_2$ is bounded by $\sqrt{n}$, due to the boundedness of the cost function. The second one is due to $\nabla_{\pi_i} \Phi(\pi) = \nabla_{\pi_i} c_i (\pi)$ and $\|c_i(\cdot, \pi_{-i})\|_2 \leq m$. The third one is by noticing $\Phi(\pi)$ is linear in $\pi_i$ for any $i \in [n]$.
\end{proof}
\newpage

\newpage
\section{Analysis for Markov potential game}

\subsection{Proof of Theorem \ref{thm:nash_mpg}}
\nashregretmpg*

\begin{proof}
    Denote $\pi_i^{\prime,t} = \min_{\pi_i} V_i^{\pi_i, \tilde{\pi}_{-i}^t}(\mu)$, we first decompose the regret as 
    \begin{align*}
         \text{Nash-regret} (T) 
         = \ & \sum^T_{t=1} \max_{i \in [n]} \left(V_i^{\tilde{\pi}_i^t, \tilde{\pi}_{-i}^t}(\mu) - \min_{\pi_i} V_i^{\pi_i, \tilde{\pi}_{-i}^t}(\mu)\right) \\
         = \ & \sum^n_{i=1} \sum^T_{t=1} \left(V_i^{\tilde{\pi}_i^t, \tilde{\pi}_{-i}^t}(\mu) - V_i^{\pi_i^t, \pi_{-i}^t}(\mu)\right) +\sum^n_{i=1} \sum^T_{t=1}  \left(V_i^{\pi_i^{\prime,t}, \pi_{-i}^t}(\mu) - V_i^{\pi_i^{\prime,t}, \tilde{\pi}_{-i}^t}(\mu)\right) \\
         & \ + \sum^n_{i=1} \sum^T_{t=1} \left(V_i^{\pi_i^t, \pi_{-i}^t}(\mu) - V_i^{\pi_i^{\prime,t}, \tilde{\pi}_{-i}^t}(\mu)\right) \\
         \leq \ & \frac{n}{\kappa}\sum^n_{i=1} \sum^T_{t=1} \|\Tilde{\pi}^t - \pi^t\|_2+ \sum^n_{i=1} \sum^T_{t=1} \left(V_i^{\pi_i^t, \pi_{-i}^t}(\mu) - V_i^{\pi_i^{\prime,t}, \tilde{\pi}_{-i}^t}(\mu)\right)\\
         \leq \ & \frac{n\mu}{\kappa}\sum^n_{i=1} \sum^T_{t=1} \|s - \pi^t\|_2+ \sum^n_{i=1} \sum^T_{t=1} \left(V_i^{\pi_i^t, \pi_{-i}^t}(\mu) - V_i^{\pi_i^{\prime,t}, \tilde{\pi}_{-i}^t}(\mu)\right)\\
         \leq \ &  \frac{\mu n^{3/2}\sqrt{S}T}{\kappa}+\sum^n_{i=1} \sum^T_{t=1} \left(V_i^{\pi_i^t, \pi_{-i}^t}(\mu) - V_i^{\pi_i^{\prime,t}, \tilde{\pi}_{-i}^t}(\mu)\right)\,,
    \end{align*}
    where the first inequality is by Lemma 18 of \cite{ding2022independent}, and the last inequality is by $\|s - \pi^t\|_2 \leq \sqrt{nS}$.

    Similar to the proof of Theorem 4.1, and by Lemma \ref{lem:dual_gap_mpg}, we have
    \begin{align*}
        \max_{\pi_i} \left(V_i^{\pi_i^t, \pi_{-i}^t}(\mu) - V_i^{\pi_i, \tilde{\pi}_{-i}^t}(\mu)\right) 
        \leq \ & \frac{1}{\kappa}\left\|\frac{d^{\pi_i^\ast, \pi_{-i}}(\mu)}{\mu}\right\|_{\infty} \max _{\pi_i^{\prime}}\left(\pi^{\prime}_i-\pi_i\right)^{\top} \nabla_{\pi_i} \Phi^\pi(\mu)\\
        \leq \ & \frac{1}{\kappa}\left\|\frac{d^{\pi_i^\ast, \pi_{-i}}(\mu)}{\mu}\right\|_{\infty} G(\pi^t)\,,
    \end{align*}
    where $G(\pi^t) = \max_\pi (\pi^t - \pi)^\top \nabla \Phi_\mu(\pi^t)$. 

    We bound $ \sum^T_{t=1} \EE \left[ G(\pi^t) \right]$ in a similar fashion as Theorem 4.1. 

    Notice that for any $\pi, \pi^\prime$, we have
    \begin{align*}
        \EE[\Phi(\pi)] 
        \leq \ & \EE\left[(H+1)\right] \leq \sum^\infty_{h = 0} (1-\kappa)^h \kappa (h+1) \leq \frac{\kappa}{1-\kappa} \sum^\infty_{h = 0} (1-\kappa)^h  h = \frac{1}{\kappa} \,,
    \end{align*}
    
    Therefore, take $\eta_t =  \frac{\kappa}{n^{3/2}mt^{3\alpha/2}} $, $\rho_t = \frac{4 \mu^{1/3}}{n^{1/3}m^{2/3} t^\alpha }$ and $\mu = \min\{\frac{1}{m^2n}, \frac{\kappa^{4/3}}{n^{7/8}m^{1/4}T^{\beta}}\}$, and by Lemma \ref{lem:est_diff_mpg} we have
    \begin{align*}
        \sum^T_{t=1} \EE \left[ G(\pi^t) \right]
        \leq \ & \frac{2}{\kappa\eta_{T+1}} + \sqrt{n S}\sum^{T}_{t=1} \left\|\nabla \Phi_\mu(\pi^t)  - d^t\right\|_2 +  \frac{2n S (1-\kappa)  }{ \kappa^3}\sum^{T}_{t=1}\eta_t \\
        \leq \ &  O \left(\frac{ n^{3/2}mT^{3\alpha/2}}{\kappa^2} + \frac{n^{3/2}mS T^{1-\beta}}{\kappa^{5/3}}  + \frac{n^{5/8}m^{3/4} S T^{1-\alpha/2+\beta/3}}{\kappa^{25/9}}  + \frac{n S T^{1-3\alpha/2}}{ \kappa^{2}} \right) \\
    \end{align*}

    Take $\alpha = 8/15$, $\beta = 1/5$, we have
    \begin{align*}
        \sum^T_{t=1} \EE \left[ G(\pi^t) \right] \leq \ & O \left(\frac{n^{3/2}mT^{4/5}}{\kappa^{2}} + \frac{n^{5/8}m^{3/4}ST^{4/5}}{\kappa^{25/9}}\right) \,.
    \end{align*}

    Combine everything, we have
    \begin{align*}
        \EE\left[\text{Nash-regret}(T)\right]
        \leq \ & O \left(\frac{ n^{3/2}mD_\infty T^{4/5}}{\kappa^{3}} + \frac{n^{5/8}m^{3/4}SD_\infty T^{4/5}}{\kappa^{25/9}}\right) \,,
    \end{align*}
    where $D_\infty = \max_i \left\|\frac{d^{\pi_i^\ast, \pi_{-i}}(\mu)}{\mu}\right\|_{\infty}$.
    
\end{proof}

\newpage
\subsection{Proof of Lemma \ref{lem:est_diff_mpg}}
\begin{restatable}[]{lem}{estdiffmpg}\label{lem:est_diff_mpg}
With $\eta_t =  \frac{1}{n^{3/2}mt^{3\alpha/2}} $, $\rho_t = \frac{4 \mu^{1/3}n^{1/3}m^{2/3} }{\kappa^{2/3}t^\alpha }$ and $\mu \geq \frac{1}{m^2n}$, we have
    \begin{align*}
        \EE \left[\|\nabla \Phi_{\mu}(\pi^t)- d^t\|_2\right] \leq O \left(\frac{n^{3/2}m\mu \sqrt{S} (1-\kappa)}{\kappa^3}  + \frac{n^{1/3}m^{2/3} \sqrt{S}}{\mu^{1/3}\kappa^{7/3}(t+1)^{\alpha/2}} \right)\,,
    \end{align*}
\end{restatable}
\begin{proof}
    First notice that
    \begin{align*}
        \EE\left[\nabla\Phi(\Tilde{\pi}^{t})\right] \leq \EE[(H+1)] \leq \sum^\infty_{h = 0} (1-\kappa)^h \kappa (h+1) \leq \frac{\kappa}{1-\kappa} \sum^\infty_{h = 0} (1-\kappa)^h  h = \frac{1}{\kappa} \,.
    \end{align*}
    
    The proof is similar to Lemma \ref{lem:est_diff}. Similarly let $a_t= \EE\left[\left\|\nabla\Phi_\mu(\Tilde{\pi}^{t}) - d^t \right\|_2^2\right]$ and take $\eta_t =  \frac{\kappa}{n^{3/2}mt^{3\alpha/2}} $, we have
    \begin{align*}
    a_t
    \leq \ & (1-\rho_t/2)a_{t-1} + \rho_t^2\EE\left[\left\|\nabla\Phi(\Tilde{\pi}^{t}) - \widehat{\nabla}_{\pi_i^t}\right\|_2^2\right]+ (1 + 2/\rho_t) \EE\left[\left\|\nabla\Phi(\Tilde{\pi}^{t}) - \nabla\Phi(\Tilde{\pi}^{t-1}) \right\|_2^2 \right] \\
    \leq \ &  (1-\rho_t/2)a_{t-1} + \rho_t^2\left(\EE\left[\left\|\nabla\Phi(\Tilde{\pi}^{t})\right\|_2^2\right] +\EE\left[\left\|\widehat{\nabla}_{\pi_i^t}\right\|_2^2\right]\right) + (1 + 2/\rho_t) \EE\left[\left\|\nabla\Phi(\Tilde{\pi}^{t}) - \nabla\Phi(\Tilde{\pi}^{t-1}) \right\|_2^2 \right] \\
    \leq \ & (1-\rho_t/2)a_{t-1} + \rho_t^2 \left(\frac{1}{\kappa^2} + \frac{24nm^2 }{\mu \kappa^4}\right)+ (1 + 2/\rho_t) \EE\left[\left\|\nabla\Phi(\Tilde{\pi}^{t}) - \nabla\Phi(\Tilde{\pi}^{t-1}) \right\|_2^2 \right] \\
    \leq \ & (1-\rho_t/2) a_{t-1} +  \frac{25nm^2 \rho_t^2 }{\mu \kappa^4} + \frac{16n^2m^2(1-\kappa)^2(1 + 2/\rho_t) }{\kappa^6}\EE\left[\left\| \Tilde{\pi}^t - \Tilde{\pi}^{t-1}\right\|_2^2 \right] \\
    \leq \ & (1-\rho_t/2) a_{t-1} + \frac{ 25nm^2 \rho_t^2 }{\mu \kappa^4} + \frac{36n^3 S m^2 \eta_t^2}{\kappa^6 \rho_t}\\
    \leq \ & (1-\rho_t/2) a_{t-1} + \frac{ 25nm^2 \rho_t^2 }{\mu \kappa^4} + \frac{36 S }{\kappa^4\rho_t}
\end{align*}
where the second inequality is by Lemma \ref{lem:grad_bound_mpg} and the definition of the potential function. 

Take $\rho_t = \frac{4n^{1/3}m^{2/3}  \mu^{1/3}}{t^\alpha }$. We want to show $a_t = \frac{C}{(t + 1)^{\alpha} } $, $C = \max\left\{a_1 2^{\alpha} , \frac{61n^{5/3} m^{7/3}S}{\mu^{2/3}\kappa^{4}}\right\}$. We show this by induction. 

The base case $t = 1$ clearly holds due to our choice of $C_1$. 

Assume the $a_{\ell-1} \leq \frac{C}{\ell^{\alpha}}$, $2 \leq \ell \leq T$, and notice that as $n,m \geq 1$, $\mu \leq 1$, we have
\begin{align*}
    a_\ell
     \leq \ & \left(1 - \frac{2n^{1/3}m^{2/3}  \mu^{1/3}}{ t^\alpha }\right) \frac{C}{\ell^\alpha} + \frac{61n^{5/3}m^{7/3}S}{\mu^{1/3}\kappa^{4} \ell^{2\alpha}} \\
     \leq \ & C \left(\frac{\ell^\alpha - 1}{\ell^{2\alpha}}\right) \\
     \leq \ & \left(\frac{C}{\ell^{\alpha}+ 1}\right)  \leq  \left(\frac{C}{(\ell+1)^{\alpha}}\right)  \,.
\end{align*}

Therefore 
\begin{align*}
\EE \left[\|\nabla \Phi_\mu(\pi^t) - d^t \|_2 \right] 
\leq \ & \EE \left[\|\nabla \Phi_\mu(\pi^t) - \nabla \Phi_\mu(\tilde{\pi}^t) \|_2 \right] + \EE \left[\|\nabla \Phi_\mu(\tilde{\pi}^t) - d^t \|_2 \right] \\
\leq \ & \frac{4nm(1-\kappa)}{\kappa^3} \EE \left[\|\pi^t - \tilde{\pi}^t\|_2\right] + \EE \left[\|\nabla \Phi_\mu(\tilde{\pi}^t) - d^t \|_2 \right] \\
\leq \ & \frac{4nm\mu (1-\kappa)}{\kappa^3} \EE \left[\|\pi^t - s\|_2\right] + \EE \left[\|\nabla \Phi_\mu(\tilde{\pi}^t) - d^t \|_2 \right] \\
\leq \ &  O \left(\frac{n^{3/2}m\mu \sqrt{S} (1-\kappa)}{\kappa^3}  + \frac{n^{5/6}m^{7/6} \sqrt{S}}{\mu^{1/3}\kappa^{2}(t+1)^{\alpha/2}} \right)\,.
\end{align*}
\end{proof}

\subsection{Proof of Theorem \ref{thm:individual_mpg}}
\individualmpg*

\begin{proof}
    Denote $\pi^\ast = \argmin_\pi \sum^T_{t=1} V_i^{\pi_i, \pi_{-i}^t}(\mu)$ by the definition of the potential function, we have
\begin{align*}
    \sum^T_{t=1} \left(V_i^{\tilde{\pi}_i^t, \tilde{\pi}_{-i}^t}(\mu) -  V_i^{\pi_i^\ast, \tilde{\pi}_{-i}^t}(\mu)\right)  
    = \ & \sum^T_{t=1} \Phi_\mu(\tilde{\pi}_i^t, \tilde{\pi}_{-i}^t) -  \Phi_\mu(\pi_i^\ast, \tilde{\pi}_{-i}^t) \\
    = \ & \sum^T_{t=1} \left(\Phi_\mu(\tilde{\pi}_i^t, \tilde{\pi}_{-i}^t) -  \Phi_\mu(\pi_i^t, \tilde{\pi}_{-i}^t) \right) + \left(\Phi(\pi_i^t, \tilde{\pi}_{-i}^t) -  \Phi_\mu(\pi_i^\ast, \tilde{\pi}_{-i}^t)\right) \\
    \leq \ & \sum^T_{t=1} \|\tilde{\pi}_i^t - \pi_i^t\| + \left(\Phi_\mu(\pi_i^t, \tilde{\pi}_{-i}^t) -  \Phi_\mu(\pi_i^\ast, \tilde{\pi}_{-i}^t)\right) \\
    \leq \ & \frac{4m\mu \sqrt{nS} T}{\kappa^3} + \left(\Phi_\mu(\pi_i^t, \tilde{\pi}_{-i}^t) -  \Phi_\mu(\pi_i^\ast, \tilde{\pi}_{-i}^t)\right) \,,
\end{align*}
where the last inequality is by the update rule and $\|\pi_i^t - s\|_2\leq \sqrt{S}$. 

Similar to the proof of Theorem \ref{thm:nash_regret}, by the smoothness of $\Phi$, $\eta_t =  \frac{\kappa}{n^{3/2}mt^{3\alpha/2}} $, $\rho_t = \frac{4 \mu^{1/3}n^{1/3}m^{2/3} }{\kappa^{2/3}t^\alpha }$ and $\mu = \min\{\frac{1}{m^2n}, \frac{\kappa^{4/3}}{n^{7/8}m^{1/4}T^{\beta}}\}$, and $\alpha = 8/15, \beta = 1/5$ we have
\begin{align*}
    & \Phi_\mu(\pi_i^{t+1}, \tilde{\pi}_{-i}^t) -  \Phi_\mu(\pi_i^\ast, \tilde{\pi}_{-i}^t) \\
    \leq \ & (1 - \eta_t) \left(  \Phi_\mu(\pi_i^{t}, \tilde{\pi}_{-i}^t) -  \Phi_\mu(\pi_i^\ast, \tilde{\pi}_{-i}^t)\right) + 2 \eta_t \sqrt{S}\| \nabla_{\pi_i} \Phi_\mu(\pi_i^{t}, \tilde{\pi}_{-i}^t) - d_i^t\| + \frac{8\eta_t^2 n^{3/2} m \sqrt{S} (1-\kappa)}{2 \kappa^3} \\
    \leq \ &  \left(1 - \frac{\kappa}{n^{3/2}mt^{3\alpha/2}}\right) \left(  \Phi_\mu(\pi_i^{t}, \tilde{\pi}_{-i}^t) -  \Phi_\mu(\pi_i^\ast, \tilde{\pi}_{-i}^t)\right) + O \left(\frac{S (1-\kappa)}{t^{3\alpha/2+\beta}\kappa^3}  + \frac{n^{27/24}m^{5/4} S}{\kappa^{7/3}t^{2\alpha-\beta/3}} + \frac{\sqrt{S}}{\kappa t^{3\alpha}}\right) \\
    \leq \ &   \left(1 - \frac{\kappa}{n^{3/2}mt^{3\alpha/2}}\right) \left(  \Phi_\mu(\pi_i^{t}, \tilde{\pi}_{-i}^t) -  \Phi_\mu(\pi_i^\ast, \tilde{\pi}_{-i}^t)\right) + O \left(\frac{n^{27/24} m^{5/4}S}{\kappa^{7/3}t}\right)\,.
\end{align*}

Using the same induction proof as the proof of Theorem 4.2, we have
\begin{align*}
     \Phi_\mu(\pi_i^{t+1}, \tilde{\pi}_{-i}^t) -  \Phi_\mu(\pi_i^\ast, \tilde{\pi}_{-i}^t) \leq \frac{D}{(t+1)^{1/5}}\,,
\end{align*}
where $D = \max \{D_1,\frac{m^{4/3}n^{27/24}S}{\kappa^{7/3}}\}$, and $D_1$ are absolute constants. 
Therefore, we have
\begin{align*}
    \sum^T_{t=1} \left(V_i^{\tilde{\pi}_i^t, \tilde{\pi}_{-i}^t}(\mu) -  V_i^{\pi_i^\ast, \tilde{\pi}_{-i}^t}(\mu)\right) \leq O \left(\frac{m^{4/3}n^{27/24}ST^{4/5}}{\kappa^{7/3}}\right)\,.
\end{align*}
\end{proof}

\newpage
\subsection{Axuiliary Lemmas}
\begin{lem}[{\cite[Lemma 2]{daskalakis2020independent}}]\label{lem:grad_bound_mpg}
    Let $\EE_t$ denote the conditional expectation on all history up to time $t$, then 
    \begin{enumerate}
        \item $\EE_t\left[\widehat{\nabla}_{\pi_i^t}\right] = \nabla_{\pi_i}\Phi^{\pi_i^t}(\mu)$.
        \item $\EE_t\left[\left\|\widehat{\nabla}_{\pi_i^t}\right\|_2^2\right] \leq \frac{24m^2}{\mu \kappa^4}$.
    \end{enumerate}
\end{lem}

\begin{lem}[{\cite[Lemma 4]{daskalakis2020independent}}]\label{lem:smoothness_mpg}
    For any $\pi, \pi^\prime$, $\left\|\nabla \Phi_{\mu}(\pi) - \nabla\Phi_{\mu}(\pi^\prime) \right\|_2 \leq \frac{4nm (1-\kappa)}{\kappa^3}\|\pi - \pi^\prime\|_2$.
\end{lem}

\begin{lem}[{\cite[Lemma D.3]{leonardos2022global}}]
\label{lem:dual_gap_mpg}
Fix any player $i$, let $\pi_i^\ast$ be an optimal policy for agent $i$ in the single agent MDP in which the rest of the agents are fixed to choose $\pi_{-i}$. Then,
\begin{align*}
    \Phi^{\pi_i^\ast, \pi_{-i}}(\mu)-\Phi^\pi(\mu) \leq \frac{1}{\kappa}\left\|\frac{d^{\pi_i^\ast, \pi_{-i}}(\mu)}{\mu}\right\|_{\infty} \max _{\pi_i^{\prime}}\left(\pi^{\prime}_i-\pi_i\right)^{\top} \nabla_{\pi_i} \Phi^\pi(\mu)\,,
\end{align*}
where $d^\pi(\mu) = \EE\left[\sum_{h=0}^{H} P^\pi\left(s^h=s \mid s_0\right)\right]$.
\end{lem}

\end{document}